\documentclass[10pt, conference]{IEEEtran}

\usepackage{float, enumerate, xcolor, algorithm, algorithmic, amsmath, graphicx, multirow, dblfloatfix, listings, multirow, tikz, amssymb, amsfonts, amsmath, amssymb, booktabs, subfigure, enumitem, authblk} 

\usepackage{tabularx}
\usepackage{makecell}

\newcommand{\ie}{{i.e.,~}}
\newcommand{\eg}{{e.g.,~}}
\newcommand{\Eg}{{E.g.,~}}
\newcommand{\etc}{{etc.}}
\newcommand{\etal}{{et~al.}} 

\newcommand{\x}{\mathbf{x}}
\newcommand{\z}{\mathbf{z}}
\newcommand{\gaussian}{\mathcal{N}}

\newcommand{\sota}{the state-of-the-art }

\newcommand{\wc}{$\ast$}

\newcommand\myeq{\mkern1.5mu{=}\mkern1.5mu}

\newcommand{\massimo}[1]{\textcolor{black}{#1}}

\newcommand{\new}[1]{\textcolor{black}{#1}}
\newcommand{\ankit}[1]{\textcolor{black}{#1}}

\newcommand{\keepme}[1]{#1}

\newcommand{\cd}[4]{\texttt{cov1d}[#1, #2, \textit{#3}, \textit{#4}] }
\newcommand{\rb}[2]{\texttt{ResblockBottleneck1D}[#1, batchnorm=\textit{#2}] }

\usepackage{array}
\newcolumntype{L}{>{\centering\arraybackslash}m{.5\columnwidth}|}

\graphicspath{ {./images/} }

\usepackage{url}

\usepackage{filecontents}
\begin{document}
	\title{Improving Password Guessing via\\Representation Learning
	}

\author[1,2,3]{\rm Dario Pasquini}
\author[1,4]{\rm Ankit Gangwal}
\author[1]{\rm Giuseppe Ateniese}
\author[3]{\rm Massimo Bernaschi}
\author[4]{\rm Mauro Conti}
\affil[1]{Stevens Institute of Technology, USA}
\affil[2]{Sapienza University of Rome, Italy}
\affil[3]{Institute of Applied Computing, CNR, Italy}
\affil[4]{University of Padua, Italy}
\affil[ ]{\textit{\{dpasquin, agangwal, gatenies\}@stevens.edu, massimo.bernaschi@cnr.it, conti@math.unipd.it}}

\maketitle
\footnotetext[1]{This paper appears in the proceedings of the 42nd IEEE Symposium on Security and Privacy (Oakland) S\&P 2021.}%
	
	\begin{abstract}
		Learning useful representations from unstructured data is one of the core challenges, as well as a driving force, of modern data-driven approaches. Deep learning has demonstrated the broad advantages of learning and harnessing such representations.
		\par
		In this paper, we introduce a deep generative model representation learning approach for password guessing. We show that an abstract password representation naturally \massimo{offers} compelling and versatile properties that open new directions in the extensively studied, and yet presently active, password guessing field. These properties can establish novel password generation techniques that are \massimo{neither feasible nor} practical with the existing probabilistic and non-probabilistic approaches. Based on these properties, we introduce:~(1) A general framework for conditional password guessing that can generate passwords with arbitrary biases; and (2) an Expectation Maximization-inspired framework that can dynamically adapt the estimated password distribution to match the distribution of the attacked password set. 	\end{abstract}

\section{Introduction}
\label{section:introduction}

Text-based passwords remain the most common form of authentication, as they are both easy to implement and familiar to users. However, text-based passwords are vulnerable to guessing attacks. These attacks have been extensively studied, and their analysis is still an active area of research. Modern password guessing attacks are founded on the observation that human-chosen passwords are not uniformly distributed in the password space (\ie all possible strings). This is due to the natural preference for choosing (easily-)memorable passwords that cover only a small fraction of the exponentially large password space. Consequently, real-world password distributions are typically composed of several dense zones that can be feasibly estimated by an adversary to perform password-space reduction attacks~\cite{yampolskiy2006analyzing}. Along that line, several probabilistic approaches have been proposed~\cite{fla,durmuth2015omen,pcfg}. These techniques - under different assumptions - try to directly estimate the probability distribution behind a set of observed passwords. Such estimation is then used to generate suitable guesses and perform efficient password guessing attacks.
\begin{figure}[t]
	\centering
	\includegraphics[trim = 30mm 35mm 35mm 30mm, clip, width=.75\linewidth]{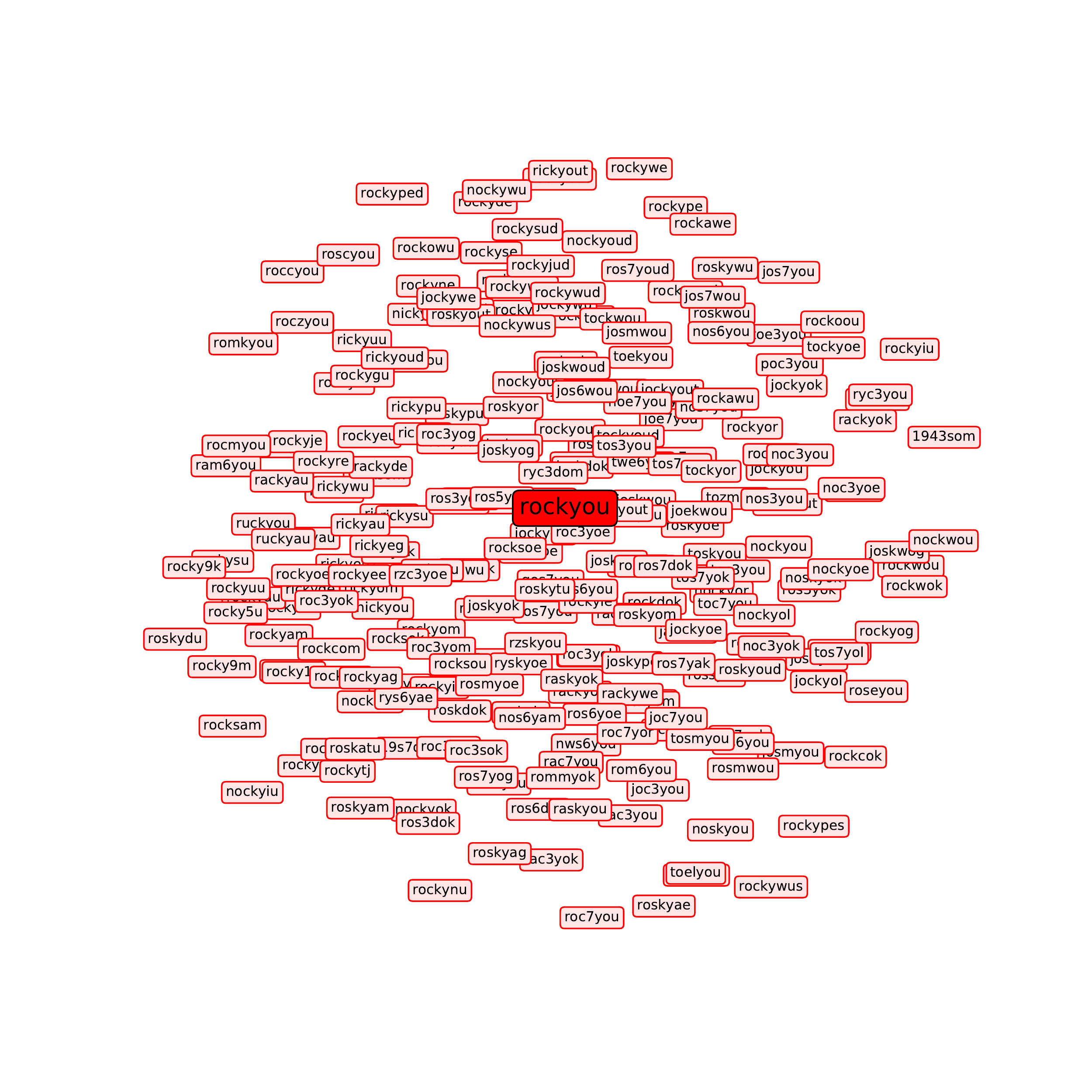}
	\vspace{-1em}
	\caption{A small section of the induced latent space around the latent point for the password ``rockyou''.}
	\label{RockYou_first}
	\vspace{-1em}
\end{figure}
\par

Orthogonal to the current lines of research, we demonstrate that an adversary can further expand the attack opportunities by leveraging representation learning techniques~\cite{bengio2013representation}. Representation learning aims at learning useful and explanatory representations~\cite{bengio2013representation} from a massive collection of unstructured data. By applying this general approach on a corpus of leaked passwords~\cite{rockyou}, we demonstrate the advantages that an adversary can gain by learning a suitable representation of the observed password distribution, rather than directly estimating it. In this paper, we show that this type of representation allows an attacker to establish novel password guessing techniques that further threaten password-based authentication systems.
\par
We model the representation of passwords in the latent space of (1) an instance of \textit{Generative Adversarial Networks}~(GANs)~\cite{goodfellow2014generative} generator and (2) an instance of \textit{Wasserstein Auto-Encoders}~(WAEs)~\cite{WAE}. This type of representation, thanks to its inherent smoothness~\cite{bengio2013representation}, enforces a semantic organization in the high-dimensional password space. Such an organization mainly implies that, in the latent space of the generator, respective representations of semantically-related passwords are closer. As a result, geometric relations in the latent space directly translate to semantic relations in the data space. A representative example of this phenomenon is loosely depicted in \figurename{~\ref{RockYou_first}}, where \massimo{we show} some latent points (with their respective plain-text passwords) localized in a small section of the induced latent space.
\par
We exploit such geometric relations to perform a peculiar form of conditional password generation. Namely, we characterize two main properties: \textit{password strong locality} and \textit{password weak locality}. These locality principles enforce different forms of passwords organization that allow us to design two novel password guessing frameworks, \textit{Conditional Password Guessing (CPG)} and \textit{Dynamic Password Guessing (DPG)}. We emphasize that \sota approaches are unable to perform such types of advanced attacks or, if somehow altered, become very inefficient. The major contributions of our work are as follows:
\begin{enumerate}
	\item We are the first to demonstrate the potential of using fully unsupervised representation learning in the password guessing domain.
	\item We introduce a probabilistic and completely unsupervised form of template-based passwords generation.
	We call this framework CPG. CPG generates arbitrarily biased passwords that can be used: (1) by an adversary to increase the impact of side channels and similar password attacks~\cite{ali2015keystroke, marquardt2011sp, vuagnoux2009compromising, balzarotti2008clearshot}; or (2) by a legitimate user to recover his/her password. We show the efficiency of~CPG \massimo{against existing solutions via experimental} evaluations.
	\item We introduce the concept of DPG: DPG is the password guessing approach that dynamically adapts the guessing strategy based on the feedback received from the interaction with the attacked passwords set. We build an Expectation Maximization-inspired DPG implementation based on the password locality enforced by the deep generative model. DPG shows that an attacker can consistently increase the impact of the attack by leveraging the passwords successfully guessed during a running attack.
\end{enumerate}
It is important to highlight that ongoing developments in deep generative frameworks would naturally translate into further improvements in our approach.
		%
\par
\textit{Organization:}
Section~\ref{section:background} gives an overview of the fundamental concepts related to our work. 
Here, we also present our model improvements and the tools upon which our core work is based. We introduce password locality along with CPG in Section~\ref{section:Passwords_Strong_locality_and_partial_knowledge_password_guessing} and DPG in Section~\ref{section:Passwords_weak_locality_and_adaptive_latent_distribution}. The evaluation of our proposed techniques is presented in their respective \massimo{sections}. Section~\ref{section:related_works} briefly discusses relevant previous works. Section~\ref{section:conclusion} concludes the paper, \massimo{although supplementary information is provided in the Appendices}.

\section{Background and preliminaries}
\label{section:background}
In Section~\ref{section:gan}, we briefly introduce deep generative models and related concepts that are important to understand our work. In Section~\ref{section:passgan}, 
we present the two deep generative models that we use as fundamental building blocks in our approach.


\subsection{Deep Generative Models}
\label{section:gan}
A deep generative model is a probabilistic model trained to perform implicit estimation of an unknown \textbf{target data distribution} $p^*(\x)$, given a set of observable data (\ie a train-set)~\cite{goodfellow2014generative, goodfellow2016nips}. In the process, a deep neural network is used to parametrize the description of the underlying data distribution.
\par
In contrast to the common \textit{prescribed probabilistic models}~\cite{diggle1984monte}, implicit probabilistic models do not explicitly estimate the probability density of data; they instead approximate the stochastic procedure that generates data~\cite{mohamed2016learning}.
\par
In the general case, deep generative models are latent variable models. That is, the network is implicitly guided to learn a set of latent variables that unfold the complex interactions among the factors describing data. During the training, a prior distribution is imposed on the learned latent variables so that we can eventually sample realizations of them after the training. Such a prior, referred to as \textbf{prior latent distribution} or $\dot{p}(\mathbf{z})$ in this paper, is an \textit{easy-to-sample}, uninformative and factorized prior. Its factorized form indicates that the network assigns a disjointed semantic meaning to each latent variable, and, consequently, learns a disentangled latent data representation for the input domain. In other words, the latent representation is modeled to capture the posterior distribution of the underlying explanatory factors of the observed data~\cite{radford2015unsupervised}. 
\par
A generative network is a deterministic mapping function $G: \mathbf{Z} \rightarrow \mathbf{X}$ between the latent space $\mathbf{Z}:\mathbb{R}^k$ and the data space $\mathbf{X}$ (\ie where the observed data is defined), specifically, \textbf{a bridge between $\dot{p}(\mathbf{z})$ and the distribution $p(\mathbf{x})$ learned by the model}. More formally, under this construction, the probabilities of data instances have the following form:
\begin{equation}
p(\mathbf{x}) = p(\mathbf{x}\mid z ; \theta)\dot{p}(\mathbf{z}),\end{equation}
where $\theta$ is the set of learnable parameters of the generator. Typical choices for $\dot{p}(\mathbf{z})$ are $\gaussian(0, \mathbf{I})$ or $U[0,1]$~\cite{goodfellow2016nips}.
\par
Sampling points $z$ from the latent space according to $\dot{p}(\mathbf{z})$ and then mapping them in the data-space through the generator, is equivalent to sampling data points from the data space $\mathbf{X}$ according to $p(\mathbf{x})$. During this operation, we can generally also consider an arbitrary $p(\z)$ that can be different\footnote{At a cost of representing a distribution different from $p^*(\x)$.} from $\dot{p}(\z)$. In the rest of this paper, we will refer to the probability density function $p(\z)$ of the latent space with the general term of \textbf{latent distribution.}
\par
Additionally, the smoothness of the generator forces a geometric organization in the learned latent space. Similar to the feature embedding techniques~\cite{goyal2018graph, li2018word}, indeed, the latent representations of semantically bounded data points show strong spatial coherence in the latent space \cite{radford2015unsupervised}.
\par 
We build our password guessing approach on top of two interchangeable deep generative model frameworks, namely, Generative Adversarial Networks and Autoencoders.
\paragraph{Generative Adversarial Networks (GANs)} 
The GANs framework learns a deep generative model by following an adversarial training approach. The training process is guided by a second network $D$ (\ie the critic/discriminator), which gives a density estimation-by-comparison~\cite{mohamed2016learning} loss function to the generative network $G$ (\ie the generator). The adversarial training bypasses the necessity of defining an explicit likelihood-function and allows us to have a good estimation of very sharp distributions~\cite{goodfellow2016nips}.\\
During the training, latent points $z$ are directly sampled from $\dot{p}(\mathbf{z})$ and given as input to $G$. In turn, the latter maps those in the data-space, where they are fed to the network $D$. The critic, receiving both ground-truth data instances from the train-set and generating data from $G$, is trained to allocate density only to real data instances. The generator $G$, instead, is adversarially trained to force $D$ to arrange probability estimates on the output of $G(z)$. The optimization follows from a coordinate minimization of the losses of the two networks.
\paragraph{Autoencoders (AE)} An Autoencoder is a deep generative model that conceptually compounds of two networks: an encoder network $Enc:\mathbf{X} \rightarrow \mathbf{Z}$ and a decoder network $Dec: \mathbf{Z} \rightarrow \mathbf{X}$. The resulting aggregate model is generally trained to learn a form of identity function: $x = Dec(Enc(x))$, or a more useful variation of it. Unlike GANs, no adversarial training is exploited during the training; typically, a maximum likelihood approach is used, instead. Once trained, the network $Dec$ can serve as a data generator where meaningful latent points are fed as input to it. However, to allow for efficient sampling from the latent space, an AE needs a form of explicit regularization during the training; that is, the latent space must be forced to be coherent with a chosen prior latent distribution. Widely known AEs implementing this strategy are described in~\cite{VAE, AAE, WAE}.
\par
In the rest of the paper, we make no distinction between the decoder network $Dec$ and the GAN generator; we refer to either of them as $G$. In the same way, we employ $E$ to refer to the encoder network used to model the inverse mapping in both  models: $G^{-1}:~\mathbf{X} \rightarrow \mathbf{Z}$. For the AE, this network is $Enc$, whereas, for the GAN,  it is the network described in the Appendix \ref{section:learn_inverse_mapping}.
 
\subsection{Password guessing with deep generative models}
\label{section:passgan}
Hitaj~\etal~in their seminal work PassGAN~\cite{PassGAN} trained a GAN generator as an implicit estimator of password distributions. 
PassGAN harnesses an off-the-shelf Wasserstein GAN with gradient penalty~\cite{IWGAN} over a residual-block-based architecture~\cite{resnet}. It assumes a latent space with a standard normal distribution as its prior latent distribution and dimensionality equal to $128$. The model is trained on the RockYou~\cite{rockyou} password leak, and only passwords with $10$ or fewer characters were considered.
Despite its underlying potential, the password guessing approach presented in PassGAN suffers from technical limitations and inherent disadvantages in its application.\footnote{As a matter of fact, PassGAN requires up to ten times more guesses to reach the same number of matched passwords as the probabilistic and non-probabilistic competitors.} Most limitations can be addressed as shown in Section \ref{section:model_improvements}. However, some limitations are intrinsic to the model itself. A prominent example is the model's inability to assign probabilities to the produced guesses consistently and thus sort them based on popularity. 
This drawback might make the GAN approach undesirable in a standard trawling scenario. However, in the present work, we show the existence of novel and valuable properties intrinsic to the class of deep generative models. Abstracting the underlying model under the perspective of representation learning, we prove that these properties can be used to devise unique guessing techniques that are infeasible with any existing approaches.
\par
Next, we introduce the necessary improvements to the original PassGAN construction (Section \ref{section:model_improvements}). In Section \ref{section:CWAE}, we introduce a different and novel deep generative model in the password guessing domain. 

\subsubsection{Improved GAN model}
\label{section:model_improvements}
The password guessing approach presented in PassGAN suffers from an inherent training instability. Under such conditions, the generator and the critic cannot carry out a sufficient number of training iterations. This may lead to an unsuitable approximation of the target data distribution and reduced accuracy in the password guessing~task. 
In the original model, the discrete representation of the strings (\ie passwords) in the train-set\footnote{Each string is represented as a binary matrix obtained by the concatenation of the one-hot encoded characters.} introduces strong instability for two main reasons:~(1)~The discrete data format is very hard to reproduce \massimo{for} the generator because of the final \textit{softmax} activation function, which can easily cause a low-quality gradient; and (2)~the inability of the generator to fully mimic the discrete nature of the train-set makes it straightforward for the critic to distinguish between real and generated data. Hence, the critic can assign the correct~\textit{``class''} easily, leaving no room for an enhancement of the generator, especially in the final stages of the training.
\par
To tackle the problems above, we apply a form of stochastic smoothing over the representation of the strings contained in the train-set. 
 This smoothing operation consists of applying an additive noise of small magnitude over the one-hot encoding representation of each character. The smoothing operation is governed by a hyper-parameter $\gamma$, which defines the upper-bound of the noise's magnitude. We empirically chose $\gamma~\myeq~0.01$ and re-normalize each distribution of characters after the application of the noise. This smoothing operation has a significant impact on the dynamics of the training, allowing us to perform $30$ times more training iterations without training collapse~\cite{brock2018large}. We keep the general GAN framework mostly unchanged because of the excellent performance of the \textit{gradient-penalty-WGAN}~\cite{IWGAN}.
\par
With our improvements in the training process, we can exploit a deeper architecture for both the generator and the critic. We substitute the plain residual blocks with deeper residual bottleneck blocks~\cite{resnet}, leaving their number intact. We find the use of batch normalization in the generator to be essential for increasing the number of layers of the networks successfully.
\par
The new architecture and the revised training process allow us to learn a better approximation of the target password distribution, and consequently, outperform the original PassGAN. A comparison between the original and our improved approach is reported in \tablename~\ref{NewVSOld}. In this experiment, both models are trained on 80\% of RockYou leak and compared in a trawling attack\footnote{Under the same configuration proposed in~\cite{PassGAN}.} on the remaining 20\% of the set. As the 20\% test-set does not contain passwords present in the train-set, the performance of a model in this test demonstrates its ability to generate new valid passwords, excluding over-fitting artifacts.
%
\begin{table}[h]
	\centering
	\caption{The matched passwords by PassGAN and our improved model over the RockYou test-set}
	\resizebox{.5\columnwidth}{!}{%
		\begin{tabular}{rrr}
			\toprule
			\multicolumn{1}{r}{\textbf{\begin{tabular}[c]{@{}c@{}}Number\\ guesses\end{tabular}}} & \multicolumn{1}{r}{\textbf{\begin{tabular}[c]{@{}c@{}}PassGAN\\ (\%)\end{tabular}}} & \multicolumn{1}{r}{\textbf{\begin{tabular}[c]{@{}c@{}}Our GAN\\ (\%)\end{tabular}}} \\
			\midrule
			$1 \cdot 10^{8}$ & 6.72 & 9.51 \\
			$1 \cdot 10^{9}$ & 15.09 & 23.33 \\
			$1 \cdot 10^{10}$ & 26.03 & 40.48 \\
			$2 \cdot 10 ^ {10}$ & 29.54 & 45.55 \\
			$3 \cdot 10 ^ {10}$ & 31.60 & 48.40 \\
			$4 \cdot 10 ^ {10}$ & 33.05 & 50.34 \\
			$5 \cdot 10 ^ {10}$ & 34.19 & 51.80 \\
			\bottomrule
		\end{tabular}
	}
	\label{NewVSOld}
\end{table}
In this work, we use the improved settings described in the present section. We train three different generators, using a 80-20\% split of RockYou leak, considering passwords with a maximum length of $10$, $16$, and $22$, respectively.
\subsubsection{Autoencoder for password guessing}
\label{section:CWAE}
To highlight the generality of the proposed approaches,~we introduce a second and novel deep generative model for password guessing. It is based on Wasserstein Autoencoder~(WAE)~\cite{WAE} with moment matching regularization applied to the latent space (called WAE-MMD~\cite{WAE}). To allow for sampling from the latent space, WAE regularizes the latent space to make it coherent with a chosen prior latent distribution.
\par
A WAE learns a latent representation that shares several properties with the one coming from the GAN-based technique. 
Nevertheless, these models naturally provide a very accurate inverse mapping, \ie $Enc$, which makes the model superior to the default GAN-based one in certain scenarios.
\par
To add further regulation to the WAE, we train the model as a Context AE (CAE)~\cite{CAE}. During every iteration of the training process of a CAE, the encoder receives a noisy version $\tilde{x}_i$ of the input password $x_i$. The noisy input is obtained by removing each of the characters in the password $x$ with a certain probability $p=\frac{\epsilon}{|x_i|}$ where $|x_i|$ is the number of characters in the password, and $\epsilon$ is a hyper-parameter fixed to $5$ in our setup. 
Our model receives the mangled input $\tilde{x}_i$, and then it is trained to reproduce the complete password as the output ($x = Dec(Enc(\tilde{x}))$); that is, the model must estimate the missing characters from the context given by the available ones. Furthermore, the CAE training procedure allows us to contextualize the wildcard character that we will use in Section~\ref{section:password_template_inversion}. We refer to our final model as the {\em Context Wasserstein Autoencoder}, or CWAE.
\par
We set up the CWAE with a deeper version of the architecture used for the GAN generator. We use the same prior latent distribution of our GAN generator, \ie $\gaussian(0, \mathbf{I})$ with a dimension of $128$. The training process is performed over the same train-sets of the GAN. 

\section{Conditional password guessing (CPG) and Passwords strong locality} 
\label{section:Passwords_Strong_locality_and_partial_knowledge_password_guessing}
In this section, we present the first contribution of our paper, \ie the password locality concept, and its possible applications for password guessing.
In Section~\ref{section:Strong_locality}, we describe the most natural form of locality that we call \textit{password strong locality}.
 In Section~\ref{section:password_template_inversion}, we demonstrate the practical application of password locality by introducing a technique that we call ``password template inversion'' for conditional and partial knowledge passwords generation. Finally, we demonstrate the advantages that our technique offers over existing probabilistic and non-probabilistic password models.
\subsection{Password strong locality and localized sampling}
\label{section:Strong_locality}
As we briefly introduced in Section~\ref{section:gan}, the latent representation learned by the generator enforces geometric connections among latent points that share semantic relations in the data space. As a result, the latent representation maintains \textit{``similar''} instances closer. 
\par
\begin{figure}
\centering
\includegraphics[trim = 45mm 45mm 45mm 45mm, clip, width=1\linewidth]{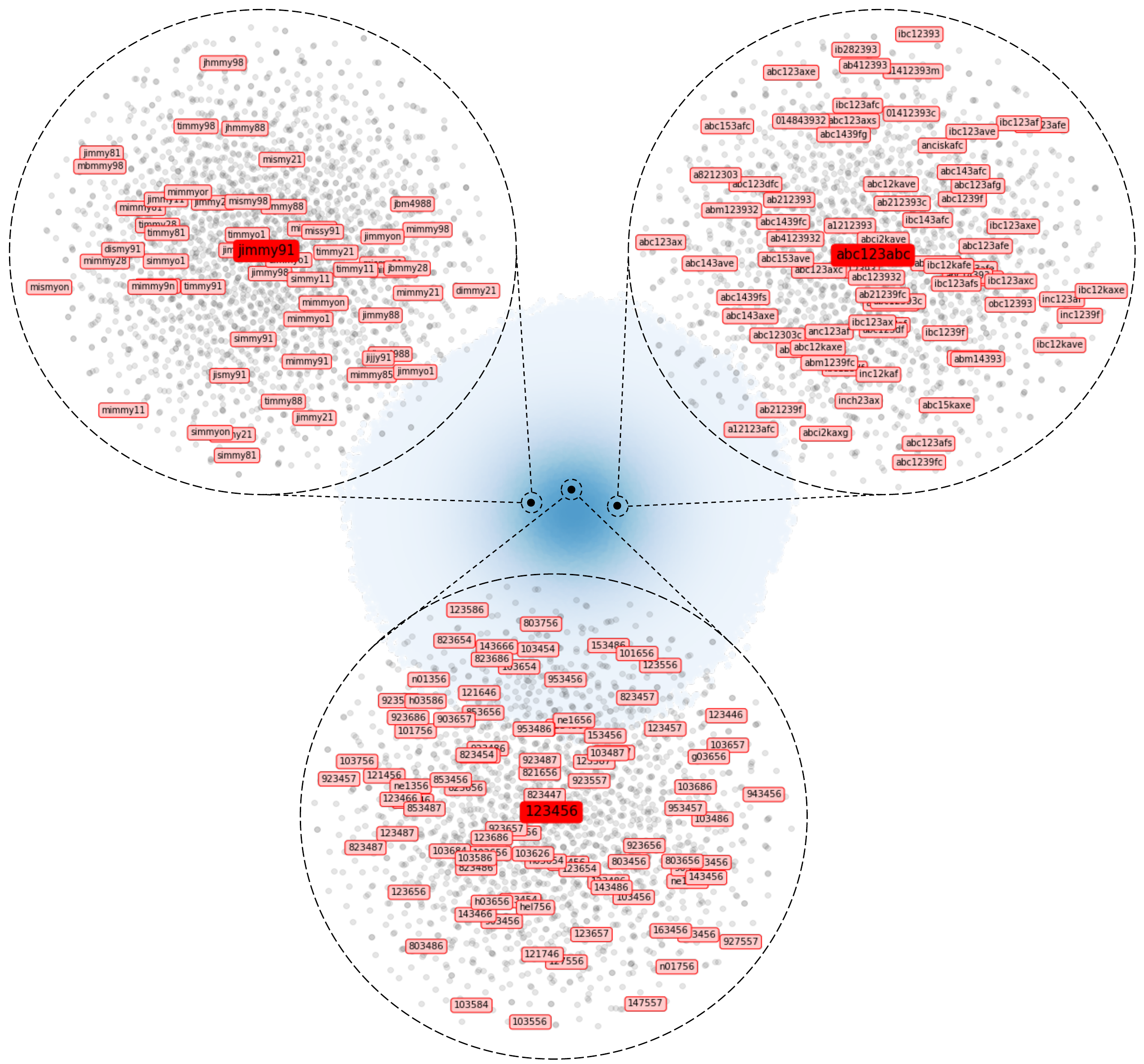}
\caption[]{2D representation of small portions around three latent points corresponding to three sample passwords ``\texttt{jimmy91}'', ``\texttt{abc123abc}'', and ``\texttt{123456}'' in the latent space learned from the RockYou train-set. Note: for the sake of better illustration, the image has been cropped.}
\label{RockYou_TSNE}
\end{figure}

\keepme{
In general, the concept of similarity harnessed in the latent space of a deep generative model solely depends on the modeled data domain (\eg images, text) and its distribution. However, external properties can be incentivized by the designer via injection of inductive bias during the training. An example is reported in Appendix \ref{sec:ind_bias}.}
 In the case of our \textbf{passwords latent representations}, the concept of similarity mainly relies on a few key factors such as the structure of the password, the occurrence of common substrings, and the class of characters. \figurename{~\ref{RockYou_TSNE}} (obtained by t-SNE~\cite{maaten2008visualizing}) depicts this observation by showing a 2D representation of small portions around three latent points (corresponding to three sample passwords ``\texttt{jimmy91}'', ``\texttt{abc123abc}'', and ``\texttt{123456}'') in the latent space. Looking at the area with password ``\texttt{jimmy91}'' as the center, we can observe how the surrounding passwords share the same general structure (5L2D,~\ie 5 letters followed by 2 digits) and tend to maintain the substring ``jimmy'' with minor variations. Likewise, the area with the string ``\texttt{abc123abc}'' exhibits a similar phenomenon, where such a string is not present in the selected train-set and does not represent a common password template.
\par
We loosely name \textbf{password strong locality} the representation's inherent property of grouping together passwords that share very fine-grained characteristics. The password strong locality property asserts that latent representation of passwords sharing some specific characteristics, such as identical substrings and structure, are organized in close proximity to each other. In Section~\ref{section:weak_locality}, we will show that strong locality also implies a weaker but general form of semantic bounding, which we refer to as weak locality.
\par
The strong locality becomes particularly compelling when selecting where to focus on the sampling operation during the password generation process. \textbf{Indeed, since different classes of passwords are organized and bounded into different zones of the underlying space, it is possible to generate specific classes of passwords by sampling from specific areas of it.}  
We leverage this technique to induce arbitrary biases in the generation process.\footnote{From the model's point of view, this is equivalent to changing the latent distribution, and in particular, reallocating its expected value to a different zone of the latent space.} However, we must first define a meaningful and practical way to express such biases, that is, to localize the zones of the latent space we are interested in. 

\par
One naive solution resorts to a prototype password $x$ to guide the localization process. In particular, we can generate passwords strictly related to the chosen prototype password $x$, by fetching latent points around the latent representation $z$ of $x$ (\ie $x = G(z)$). Thanks to the strong locality, the obtained latent points should be valid latent representations of passwords with an arbitrary strong relation with $x$. In this context, we refer to the chosen $x$ (or its corresponding latent representation $z$) with the term \textbf{pivot}. The three dark red boxes in \figurename{~\ref{RockYou_TSNE}} are the pivot points in the latent space for their corresponding passwords.
\par
To infer the latent representation $z$ from $x$, we use the encoder network described in Section~\ref{section:CWAE}, as that $z~\myeq~E(x)$. We highlight that, being this process general and model-independent, other deep generative models such as~\cite{donahue2016adversarial, VAE, AAE} can be used as well. 
\par
Once we obtain the intended pivot $z$, we can easily generate coherent passwords by restricting the generator's sampling in a confined area of the latent space around $z$ (loosely represented by the small dashed circles in~\figurename{~\ref{RockYou_TSNE}}). To that purpose, we consider a new latent distribution for the generator. The new distribution has the latent representation of the pivot password as its expected value and an arbitrarily small scale. To remain coherent with prior latent distribution and partially avoiding distribution mismatch for the sampled points~\cite{white2016sampling}, we chose a Gaussian distribution: $\gaussian(z, \sigma \mathbf{I})$.
\par
According to the concept of password locality, the strength of the semantic relation between a sampled latent point and its pivot should be proportional to the spatial distance between them. Consequently, the chosen value of $\sigma$ (\ie standard deviation) offers us a direct way to control the level of semantic bounding existing in the generated passwords. This intuition is better explained by~\tablename{~\ref{table:jimmy91}}, where passwords obtained with different values of $\sigma$ for the same pivot password are reported.
\begin{table}[t]
\centering
\caption{The first-ten passwords obtained with different values of $\sigma$ \massimo{starting from} the pivot string \textit{``jimmy91''}
}
\resizebox{.75\columnwidth}{!}{%
\begin{tabular}{cccc}
	\toprule
	$\sigma\textbf{=0.05}$ & $\sigma\textbf{=0.08}$ & $\sigma\textbf{=0.10}$ & $\sigma\textbf{=0.15}$ \\ 
	\midrule
	jimmy91 & jimmy99 & mnmm988 & jimmy91992 \\
	jimmy11 & micmy91 & tbmmy98 & jrm6998 \\
	jimmy21 & jimsy91 & jismyo15 & sirsy91 \\
	jimmy88 & mimmyo1 & jizmyon & jrz4988 \\
	jimmy81 & jbmmy88 & j144988 & Rimky28 \\
	jimmy98 & simmy98 & jbmm998 & missy11 \\
	mimmy98 & dijmy91 & timsy91 & jimmy119 \\
	jimmy28 & jimmy98 & jrm4985 & sikjy91 \\
	simmy91 & timsy91 & jhmmy88 & licky916 \\
	mimmy91 & jnmm988 & jhmm988 & gimjyon \\ 
	\bottomrule
\end{tabular}
}
\label{table:jimmy91}
\end{table}
\begin{table*}[!b]
\centering
\caption[]{An example of exploiting strong locality property over a generator trained on RockYou train-set for some password templates. \new{Passwords are generated by sampling $10000$ strings with $\alpha = 0.8$ and reported in decreasing frequency order.}}
\resizebox{1.9\columnwidth}{!}{%
\begin{tabular}{lllllllllll}
	\toprule
	\multicolumn{1}{c}{\textbf{A}} & \multicolumn{1}{c}{\textbf{B}} & \multicolumn{1}{c}{\textbf{C}} & \multicolumn{1}{c}{\textbf{D}} & \multicolumn{1}{c}{\textbf{E}} & \multicolumn{1}{c}{\textbf{F}} & \multicolumn{1}{c}{\textbf{G}} & \multicolumn{1}{c}{\textbf{H}} & \multicolumn{1}{c}{\textbf{I}} & \multicolumn{1}{c}{\textbf{L}} & \multicolumn{1}{c}{\textbf{M}} \\
	\midrule
	\textbf{jimmy\wc\wc} & \textbf{jimmy\wc\wc\wc\wc} & \textbf{\wc\wc jimmy} & \textbf{\wc\wc mm\wc91} & \textbf{\wc\wc\wc\wc\wc91} & \textbf{12\wc\wc\wc91} & \textbf{A\wc\wc\wc\wc\wc} & \textbf{\wc\wc\wc A\wc\wc\wc} & \textbf{Ra\wc\wc\wc\wc91} & \textbf{(\wc\wc\wc1\wc\wc\wc)} & \textbf{\wc\wc\wc\#\wc\wc!!!} \\
	\midrule
	jimmy11 & jimmybean & majimmy & summy91 & 1111991 & 1231991 & Andres & RONALDO & Raider91 & (2001999)& 123\#1!!!! \\
	jimmy13 & jimmybear & mujimmy & sammy91 & 9111991 & 1211991 & ANDRES & MALANIA & Rainer91 & (1701939) & tom\#!!!!! \\
	jimmy01 & jimmy1001 & mojimmy & tommy91 & a111991 & 1221991 & Andrea & MANANA1 & Rain1991 & (toe1234) & bom\#!!!!! \\
	jimmy12 & jimmyjean & myjimmy & tammy91 & jan1991 & 1201991 & A10123 & SALAN11 & Raidel91 & (13@1932) & Bom\#1!!!! \\
	jimmy10 & jimmylove & 12jimmy & mommy91 & cao1991 & 1271991 & Angela & RATALIS & Ranger91 & (gar1k()) & Bam\#99!! \\
	jimmy20 & jimmy2004 & jojimmy & jimmy91 & ban1991 & 1234591 & A12123 & 123A123 & Rana1991 & (1031123)& 190\#1!!!! \\
	jimmy21 & jimmy1234 & gojimmy & gimmy91 & 5121991 & 1219991 & Andrey & BRIANA1 & Raid1991 & (1231234)& abc\#2!!!! \\
	jimmy16 & jimmybabe & jjjimmy & iammy91 & man1991 & 1205091 & ANDREY & MALA123 & Raynay91 & (sot123)) & 123\#11!!!'\\
	jimmy19 & jimmygirl & aajimmy & mimmy91 & 1811991 & 1280791 & ABC123 & AAIANA1 & Rayder91 & (Go)12(7)& Bom\#\#!!!! \\
	jimmyes & jimmy1000 & m0jimmy & sommy91 & jao1991 & 12g1991 & ABERES & BALAND1 & RaIN1991 & (11\_199\%) & 123\#16!!!\\
	\bottomrule
\end{tabular}
}
\label{table:wildcards}
\end{table*}
Lower values of $\sigma$ produce highly aligned passwords, whereas larger values of $\sigma$ allow us to explore areas far from the pivot and produce a different type of \textit{``similar''} passwords. As shown in~\tablename{~\ref{table:jimmy91}}, all the passwords generated with $\sigma~=~0.05$ retained not only the structure of the pivot (\ie 5L2D), but also observed minor variations coherent with the underlying password distribution. 
Of note, passwords generated with $\sigma = 0.15$ tend to escape the password template imposed by the pivot and reaching related-but-dissimilar password structures (\eg ``\texttt{jimmy91992}'' and ``\texttt{j144988}''). 

\subsection{Localized passwords generation with password template inversion}
\label{section:password_template_inversion}
As briefly discussed in Section{~\ref{section:Strong_locality}}, the password locality property offers a natural way to generate a very specific/confined class of passwords for a chosen pivot, a task accomplished by exploiting an encoder network $E$. This encoder is trained to approximate the inverse function $G^{-1}$, and it is the only tool we have to explore the latent space meaningfully. The default behavior of the encoder is to take as an input a string $s$ and precisely localize the corresponding latent representation in the latent space. As shown in~\tablename{~\ref{table:jimmy91}}, sampling from a distribution centered on the obtained latent point, allows us to generate a set of related passwords. However, this approach alone is not sufficient within the password guessing scenario. 
\par
In this section, we show that it is possible to ``trick'' the encoder network into further localizing general classes of passwords. We can arbitrarily define these classes via a minimal \textbf{template}, which expresses the definition of the target password class. 
\par
The encoder network can be forced to work around a specific password definition by introducing a \textbf{wildcard} character into its alphabet. The wildcard character - represented by the symbol `\wc'~in the present paper - can be used as a placeholder to indicate an unspecified character. For instance, the template ``\texttt{jimmy\wc\wc}'' expresses a class of passwords starting with the string ``jimmy'' followed by two undefined characters. When the encoder inverts this string, the obtained latent point represents the \textbf{center} of the cluster of passwords in the latent space with a total length of $8$~characters and a prefix ``jimmy''. Therefore, sampling around this latent point allows us to generate good instantiations (according to $p(\x)$) of the input template. Column~\textit{A} of~\tablename{~\ref{table:wildcards}} shows an example for the template ``\texttt{jimmy\wc\wc}''. In practice, we implement this behavior by mapping a wildcard character to an empty one-hot encoded vector when the matrix corresponding to the input string is given to the encoder. The wildcard characters can be placed in any position to define an arbitrarily complex password template; some examples are reported in~\tablename{~\ref{table:wildcards}}.
\par
Relying on this technique, the template inversion guides us towards the most plausible zone of the latent space.
When we sample from that zone, the wildcards are replaced with high-probability characters according to the distribution $p(\x)$, \ie the probability distribution modeled by the generator. This phenomenon can be observed in the generated samples (Column~\textit{A} of~\tablename{~\ref{table:wildcards}}): wildcards in most of the generated passwords have been replaced with digits to conceivably reproduce the frequent password pattern `\textit{lower\_case\_string+digits}'~\cite{ur2015added}. On the contrary, passwords from the template ``\texttt{\wc\wc\wc\wc\wc91}'' are reported in Column~\textit{E} of~\tablename{~\ref{table:wildcards}}. In this example, we ask the generator to find 7-character long passwords where the last two characters are digits. Here, the generated passwords tend to lie towards two most likely password classes for this case, \ie `\textit{lower\_case\_string+digits}' complementary to the previous case and `\textit{all\_digits}.' As the localized zone of the latent space is a function of all the observed characters, the same template with more observable digits (\eg Column~\textit{F} of~\tablename{~\ref{table:wildcards}}) ends up generating \textit{all\_digits} passwords with higher probability. 
\subsection{Conditional Password Guessing (CPG)}
\label{section:CPG}
\keepme{
One of the most significant limitations of available probabilistic guessers is their intrinsic rigidity. The inductive bias imposed on such models allows them to be extremely suitable for general trawling attacks, yet it causes them to fail at adapting to different guessing scenarios. For instance, they fail to handle a natural as well as a general form of conditional password generation, such as the template-based one that we proposed in Section~\ref{section:password_template_inversion}. Despite the limitations of existing approaches, generating guesses under arbitrary biases is a useful and helpful procedure. This applies to both security practitioners and common users. Some examples are below:
\begin{itemize}
	\item An attacker can be interested in generating an arbitrary number of guesses having a particular structure or common substring. For instance, an attacker might want to generate passwords containing the name of the attacked web application as substring.\footnote{It has been widely observed that many users tend to incorporate such names in their passwords.} 
	\item A conditional password generation capable of working with partial knowledge can be used by an attacker to improve the impact of side-channel attacks targeting user input~\cite{ali2015keystroke, marquardt2011sp, vuagnoux2009compromising, balzarotti2008clearshot}. These attacks often recover only an incomplete password (\eg some characters) due to their accuracy. An attacker can leverage conditional password generation mechanisms to input missing characters and recover the target password.
	\item Similarly, a legitimate user can be interested in recovering her/his forgotten password while remembering a partial template, for example, ``***Jimmy**1**8\#''.
\end{itemize}
In this direction, conditional password generation is particularly difficult for \textit{autoregressive} password guessers, such as the RNN-based ones (\eg FLA~\cite{fla}). Indeed, these approaches, in the general case, are unable to assign a probability to missing characters of a template efficiently; the forward-directionality, intrinsic in their generation process, eliminates the possibility of an efficient appreciation of wildcards occurring before a given substring (\eg the case in Columns~\textit{C} and \textit{E} of~\tablename{~\ref{table:wildcards}}). In these cases, the probability of an exponential number of passwords could be computed before using the characters in the template to prune the visit tree. This is the case of the template reported in Column~\textit{E}, where the required computational cost for these approaches is not far from computing all the passwords into the chosen probability threshold and filter the ones coherent with the template. More generally, these approaches cannot be efficiently applied when a large number of wildcards is considered. Sampling from the posterior distribution over the missing variables (\ie wildcards), indeed, is intractable for not minimal alphabets; for instance, for an alphabet of size $|\Sigma|$ , it requires $O(|\Sigma|)$ runs of network inference per step of Gibbs sampling or iterated conditional modes \cite{bowman-etal-2016-generating}.
Yet, they can handle the generation for a special case of templates (\eg Column~\textit{A} and Column~\textit{B}), where the prefix of the template is fully known, and no observable character appears among the \textit{wildcards}.}
\par
To generate over arbitrary templates, a possible trivial approach for autoregressive models would be to enumerate passwords according to the chosen cut-off probability and then filter the ones compatible with the chosen bias. However, this solution has two main drawbacks. First, this operation is costly, as well as storage-demanding. More significantly, such an approach can easily become intractable for small cut-off probability values, as the enumeration could require an exponential-scale cost due to the unpruned visit of the space. The second and more substantial limitation of this approach resides in the difficulty of generating relative low-probability guesses. In other words, \textbf{if the chosen bias results in candidate passwords having low probabilities (according to the estimated password distribution), those will be unlikely generated during the enumeration process, at least, for a reasonable cut-off probability.} In turn, this translates into the impossibility of enumeration-based approaches to generate the number of valid guesses required to a sound password guessing attack. 

By contrast, conditional password generation can seamlessly be implemented within our representation-learning-based approach and its locality property. The password organization imposed by this locality principle maintains similar passwords bounded in a precise zone of the latent space. Localizing such zones using the template inversion technique and sampling from them allow us to enumerate biased passwords with minimal effort. 
We can conditionally produce suitable guesses for each meaningful bias, even if this yields low probability passwords. Algorithm~\ref{algorithm:cpg_attack} briefly formalizes this approach. Chosen a template $t$, we use the encoder network $E$ to obtain the latent representation $z^t$ of $t$. Then, we sample latent points from a distribution centered in $z_t$ and with scale $\sigma$. During the process, we filter the guesses coherent with $t$ (\textit{if} statement at line~$6$).
The effectiveness of this conditional guesses generation process will be demonstrated in the next~section.
\begin{algorithm}[!htbp]
	\caption{Conditional Password Guessing (CPG)}
	\label{algorithm:cpg_attack}
	 \scriptsize
	\begin{algorithmic}[1]
		\renewcommand{\algorithmicrequire}{\textbf{Input:}}
		\renewcommand{\algorithmicensure}{\textbf{Output:}}
		\REQUIRE Template: $t$, Int: $n$, Real: $\sigma$
		\ENSURE Passwords set: $X$
		\STATE$X=\{\}$
		\STATE$z^t = E(t)$
		\FOR{i:=1 \TO n}
		\STATE $z_i \sim \gaussian(z^t, \sigma \mathbf{I})$
		\STATE $x_i = G(z_i)$
		\IF{$x_i \vdash t$}
		\STATE $X = X \cup \{x_i\}$
		\ENDIF
		\ENDFOR
		\RETURN $X$
	\end{algorithmic}
\end{algorithm}
\subsection{Evaluation}
In this section, we evaluate our proposed CPG framework against \sota password guessers.
\subsubsection{Biased test-sets creation}
\label{sec:eval_cpg}
To create a suitable scenario to evaluate our conditional generation technique CPG, we cast a set of biased password test-sets. In our setup, a bias $t_i$ is a password template; a string $t_i \in \{ \Sigma\cup\{*\} \}^*$ where $\Sigma$ is the password alphabet ($210$ unicode characters in our case) and $*$ is the wildcard character. Every password template $t_i$ is randomly extracted from a password sampled from a validation set $X_v$. We chose the \textit{LinkedIn}~\cite{linkedin_leak} password leak as the validation-set. From this set, we keep passwords with length $16$ or less, obtaining $6\cdot 10^{7}$ unique passwords, which is $\sim5$~times the RockYou train-set used to train our~model.
\par
More precisely, sampled a ground-truth password $x$ from $X_v$, we derive $t_i$ by substituting (with a certain probability $p$) each character in $x$ with a wildcard (\eg from x=``jimmy1991'' to t=``\wc i\wc my\wc\wc\wc 1''). In our setup, we select $p=0.5$. In this process, we select only those of the produced templates that contain at least $4$ observable characters and at least $5$ wildcards. The latter constraint aims at rendering not trivial a brute-force solution ($\sim 3 \cdot 10^{11}$).
\par
After obtaining a large enough collection of valid templates, we create a set of biased password test-sets. This is achieved by collecting all the passwords matching the templates in $X_v$ with an exhaustive search. More precisely, for each template, we collect all the instances $x$ of $X_v$, such that $x$ satisfies the template $t_i$; that is, the set $X_v^{t_i}= \{x| x \in X_v \land x \vdash t_i\}$. Based on the cardinality of the various $X_v^{t_i}$, we divide those into four classes:
\begin{enumerate}
	\item $T_{\text{common}}$, if $|X_v^{t_i}| \in [1000, 15000]$
	\item $T_{\text{uncommon}}$, if $|X_v^{t_i}| \in [50, 150]$
	\item $T_{\text{rare}}$, if $|X_v^{t_i}| \in [10, 15]$
	\item $T_{\text{super-rare}}$, if $|X_v^{t_i}| \in [1, 5]$
\end{enumerate}
Eventually, each of the $4$ classes of templates composes of $30$ different template sets (\ie$X_v^{t_i}$). Samples of these templates and respective matching passwords are reported in \tablename~\ref{table:template_samples} in Appendix~\ref{section:supplementary_table_and_figures}.
\par
In the next section, we will use the created biased password sets to evaluate the proposed CPG framework with a set of probabilistic and non-probabilistic state-of-the-art password guessers. We evaluate the ability of each guesser to match the passwords contained in every biased set~$X_v^{t_i}$.
\subsubsection{Results}
We perform our guessing attack using the CWAE. This model showed slightly better performance than the GAN approach in this guessing scenario.\footnote{This is due to the higher quality of the encoder network included with the auto-encoder.} We report results for the model trained on passwords with a maximum length of $16$, as no consistently different results have been obtained with models trained on password lengths $10$ and $22$.
\par
In our setup, we follow the CPG described in Section \ref{section:CPG}. More precisely, for each biased password set $X_v^{t_i}$, we invert the template $t_i$ using the encoder network. Then we sample password around the obtained latent vector using standard-deviation $\sigma=0.8$ (see Algorithm \ref{algorithm:cpg_attack}). We generate $n=10^{7}$ valid passwords for each template, and then we compute the cardinality of the intersection of the generated guesses with $X_v^{t_i}$ to calculate the number of the guessed passwords.
\par

\keepme{
We compare our CPG with five state-of-the-art guessers; namely, OMEN~\cite{durmuth2015omen} and FLA~\cite{fla} for the fully-probabilistic, PCFG~\cite{pcfg} for token-based probabilistic, and HashCat~\cite{hashcat} for non-probabilistic class. Additionally, we compare against a \textit{min-auto} configuration~\cite{ur2015measuring}.\\
 As these guessers are not able to perform a natural form of conditional password generation, we exploit the naive approach discussed in Section~\ref{section:CPG}; that is, we generate a large number of passwords in default mode and then filter the guesses coherently with the requested bias. In particular, we produced $10^{10}$ passwords for each approach. Details on the specific setup of these tools follow:
\begin{itemize}
	\item \textbf{OMEN:} We trained the Markov chain using the same train-set used for our deep generative model (\ie $80\%$ RockYou). After that, we generated $10^{10}$ sorted guesses.
	\item \textbf{PCFG:} Like in the OMEN case, we used the train-set employed for the training of our deep generative model to infer the grammar.
	\item \textbf{HashCat:} We performed a mangling rules-based attack leveraging the train-set used for the training of our deep generative model as a dictionary (considering only unique passwords sorted by frequency), and we use PasswordsPro \cite{insidepro} as the set of rules. We chose the latter based on a suitable number of rules (\ie $3120$) that allowed us to produce a suitable number of guesses. 
	\item \textbf{FLA:} We trained the largest model described in~\cite{fla}, \ie an RNN composed of three LSTM layers of $1000$ cells each and two fully connected layers. The training is carried out on the same train-set used for our model. 
	\item \textbf{CMU-PGS:} In CMU Password Guessability Service (PGS) \cite{cmupgs}, the passwords are guessed according to the \textit{min-auto} configuration~\cite{ur2015measuring}, where guesses of multiple tools (\ie FLA, Hashcat, John The Ripper, PCFG, Markov Model) are combined. We query the guess-numbers via the web interface and consider passwords requiring fewer than $10^{10}$ guesses. Recommended tools setup and \textit{``1class1''} have been used.
\end{itemize}
}
When we test each of these guessers in the conditional generation, we transform each template in a regular expression (\ie replacing the wildcards with the point operator) and extract all the guesses matching the template in the $10^{10}$ generated passwords. 
Then, we compute the cardinality of the intersection of the correct guesses with each $X_v^{t_i}$ to explicit the number of the guessed passwords.
\par
The mean percentage of guessed passwords for each templates class is reported in \tablename{~\ref{table:results_cpg}}. Coherently with the discussion done in Section~\ref{section:CPG}, our CPG framework allows us to produce a large number of biased guesses, and it matched a large portion of passwords accordingly.
\par
\begin{table}[!htbp]
	\centering
	\caption{Average matched passwords (and relative standard deviation) over the biased passwords test-set divided into $4$ classes.}
	\label{table:results_cpg}
	\resizebox{1\columnwidth}{!}{%
		\begin{tabular}{c|ccccc|cc}
			\toprule
			\textbf{\makecell{Templates\\class}} & \textbf{OMEN} & \makecell{\textbf{HashCat}\\(PasswordPro)} & \textbf{PCFG} & \keepme{\textbf{FLA}} & \makecell{\keepme{\textbf{CMU-PGS}}\\(min-auto)} & \makecell{\textbf{Our CPG}\\(CWAE)} \\ \toprule
			
			\makecell{Common\\ {[}1000-1500{]}} & \makecell{0.4383 \\($\pm~$0.1835)} & \makecell{0.5563\\($\pm~$0.1274)} & \makecell{0.7546 \\ ($\pm~$0.092)} & \makecell{0.7936\\($\pm~$0.0757)} &
			\makecell{\textbf{0.8617}\\($\pm~$0.0517)} &
			\makecell{ 0.8136\\($\pm~$0.0641)} \\ \midrule
			
			\makecell{Uncommon\\ {[}50-150{]}} & \makecell{0.2744\\($\pm~$0.1322)} & \makecell{0.3656\\($\pm~$0.1897)} & \makecell{0.5794\\($\pm~$0.1987)} & \makecell{0.6365\\($\pm~$0.1137)} & \makecell{0.7208\\($\pm~$0.1015)} &
			\makecell{\textbf{0.8606}\\($\pm~$0.0686)} \\ \midrule
			
			\makecell{Rare\\ {[}10-15{]}} & \makecell{0.1182\\($\pm~$0.1272)} & \makecell{0.2007\\($\pm~$0.1655)} & \makecell{0.4013\\($\pm~$0.2514)} &
			\makecell{0.3983\\($\pm~$0.1827)} & \makecell{0.5102\\($\pm~$0.2005)} & \makecell{\textbf{0.8482}\\($\pm~$0.1444)} \\\midrule
			
			\makecell{Super-Rare\\ {[}1-5{]}} & \makecell{0.0555\\($\pm~$0.1448)} & \makecell{0.0900\\($\pm~$0.1700)} & \makecell{0.1527\\($\pm~$0.2298)} & 
			\makecell{0.1500\\($\pm~$0.2961)} & \makecell{0.2277\\($\pm~$0.2763)} & \makecell{\textbf{0.7722}\\($\pm~$0.2910)} \\ \bottomrule
		\end{tabular}
	}
\end{table}
\keepme{
As anticipated, CPG maintains a high match ratio (\ie $>70\%$) for each template class independently of the corresponding passwords' low probabilities.
In contrast, other guessers are not able to produce such a specific class of passwords. Therefore, they provide shallow coverage of the rare templates. This is also true for the \textit{min-auto} attack, where heterogeneous guesses from multiple tools are combined. For instance, the \textit{min-auto} approach would require three orders of magnitude more guesses to match the same number of passwords as ours in the edge-case of the \textit{Super-Rare} templates.
Interestingly, given the strong bias imposed during the generation, CPG matches most passwords of other \textit{single} guessers also under the common templates case. The second best guesser turns out to be FLA that matches a comparable number of passwords as ours in the case of common templates and matches an acceptable number of passwords in the uncommon and rare classes (\ie $\geq40\%$). 
Note that we limited our CPG to generate $10^7$ guesses per template; however, more biased passwords can be sampled in a linear cost.}


\section{Dynamic Password Guessing (DPG) and Passwords weak locality }
\label{section:Passwords_weak_locality_and_adaptive_latent_distribution}
In this section, we present our major contribution, \ie Dynamic Password Guessing. In Section~\ref{section:weak_locality}, we outline the concept of password weak locality. Section~\ref{section:dynamic_attack_in_overview} introduces DPG from theoretical (Section~\ref{section:dynamic_attack_in_theory}) as well as practical (Section~\ref{section:dynamic_attack_in_practice}) viewpoints.
\begin{figure*}[!b]
	\centering
	\resizebox{.75\linewidth}{!}{%
		\vspace{-0.25em}
		\centering
		\subfigure[\textit{myspace}]{
			\centering
			\includegraphics[trim = 35mm 35mm 35mm 35mm, clip, width=.27\linewidth]{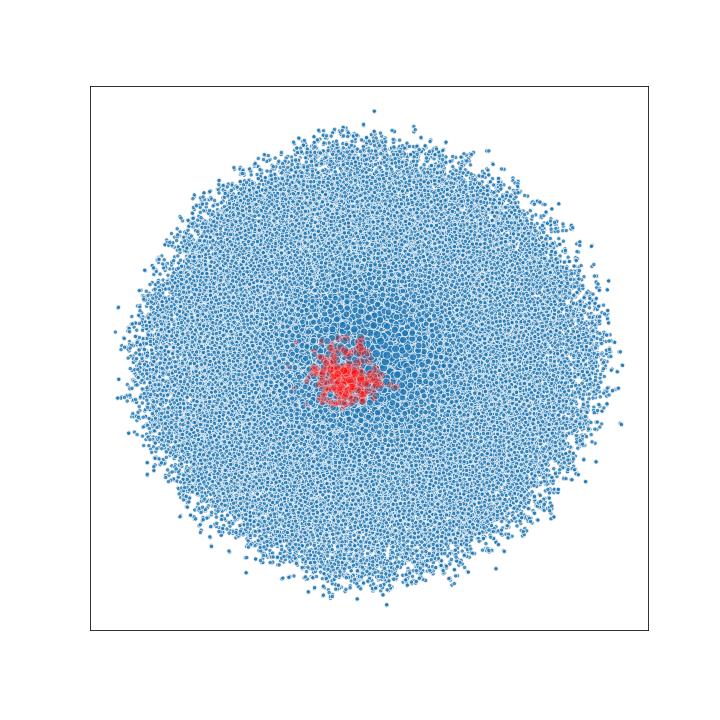}
		}\hspace{.5em}
		\subfigure[\textit{hotmail}]{
			\centering
			\includegraphics[trim = 35mm 35mm 35mm 35mm, clip, width=.27\linewidth]{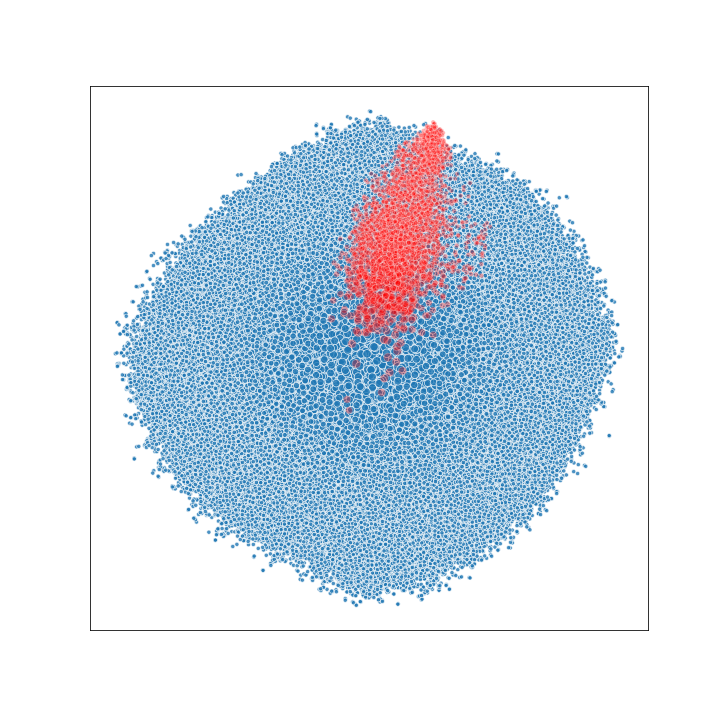}
		}\hspace{.5em}
		\subfigure[\textit{phpbb}]{
			\centering
			\includegraphics[trim = 35mm 35mm 35mm 35mm, clip,
			width=.27\linewidth]{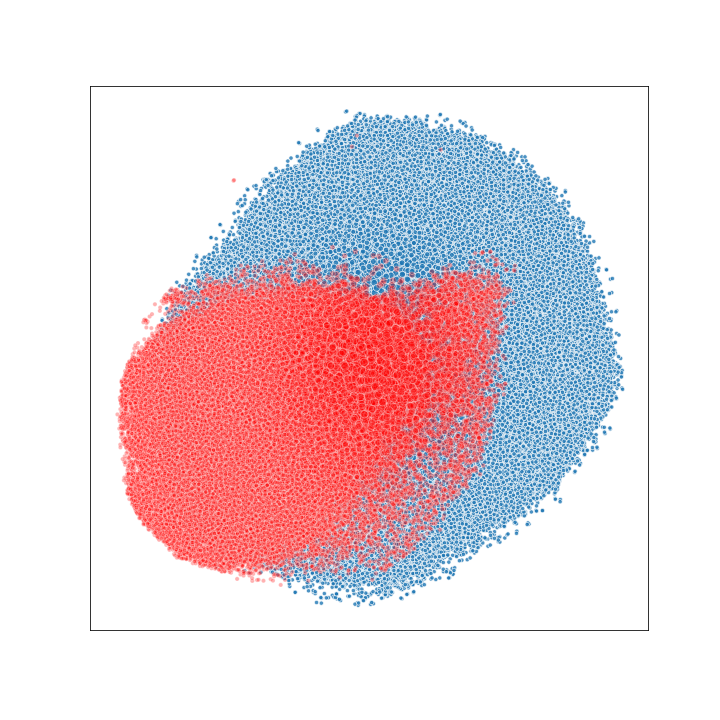}
		}
	}
	\caption[]{Password Weak Locality: 2D visualization of the latent points for three different passwords sets for a generator trained on \massimo{the} RockYou train-set. The red points represent the latent points corresponding to the passwords in the respective password set \massimo{whereas} the blue points loosely represent the dense part of the latent space. Please refer to the color version for better illustration.}
	\label{figure:dynamic_myspace}
\end{figure*}
\subsection{Password weak locality}
\label{section:weak_locality}
The embedding properties of the latent representation map passwords with similar characteristics close to each other in the latent space. We called this property strong locality, and we exploited it to generate variants of a chosen pivot password or template (discussed in Section~\ref{section:Strong_locality}). In that case, the adjective ``strong'' highlights the strict semantic relation among the generated set of passwords. However, the same dynamics enable a broader form of semantic bounding among passwords. This latter property partially captures the general features of the entire password distribution. Such features could be very abstract properties of the distribution, such as the average passwords length and character distribution ascribable to password policies. We refer to this observed property as \textbf{password weak locality} to contrast it with the strong locality.
\par 
As a representative example, \figurename{~\ref{figure:dynamic_myspace}} depicts the 2D representation of passwords from \textit{myspace}~\cite{myspace}, \textit{hotmail}~\cite{hotmail}, and \textit{phpbb}~\cite{phpbb} on the latent space learned by a generator.\footnote{It is important to emphasize that these graphical depictions are obtained by a dimension reduction algorithm. Hence, they do not depict latent space accurately. So, they merely serve as a representative illustration. We will verify our assumption empirically later in the paper.} We can observe that the passwords coming from the same dataset tend to be concentrated in the latent space and do not spread abruptly all over the spectrum. 
The dimensionality of the fraction of latent space covered by an entire password set (the red parts in~\figurename{~\ref{figure:dynamic_myspace} (a), (b), and (c)} clearly depends on the heterogeneity of its passwords. Passwords from smaller sets (\eg \textit{myspace}) are concentrated in restricted and dense zone of the latent space, whereas passwords from larger sets (\eg as \textit{phpbb}) tend to cover a more significant section while they are still tightly knitted.

\par
In the following sections, we will present evidence of this locality property, and we will show how to exploit it to improve password guessing.

\subsection{DPG for covariate shift reduction}
\label{section:dynamic_attack_in_overview}
First, we present the theoretical motivation behind DPG in Section~\ref{section:dynamic_attack_in_theory} followed by its instantiation in Section~\ref{section:dynamic_attack_in_practice}.
\subsubsection{Theoretical motivation}
\label{section:dynamic_attack_in_theory}
Probabilistic password guessing tools implicitly or explicitly attempt to capture the data distribution behind a set of observed passwords, \ie the train-set. This modeled distribution is then used to generate new and coherent guesses during a password guessing attack. A train-set is usually composed of passwords that were previously leaked. By assumption, every password-set leak is characterized by a specific password distribution $p^*(\x)$. When we train the probabilistic model, we implicitly assume $p^*(\x)$ to be general enough to well-represent the entire class of password distributions. This generality is essentially due to the fact that the real-word password guessing attacks are indeed performed over sets of passwords that potentially come from completely different password distributions. As a matter of fact, we typically do not have any information about the attack-set distribution. This can indeed be completely different from the one used for model training. As a representative example, different password policies or users' predominant languages can cause the test-set's distribution to differ from the train-set's distribution drastically. This discrepancy in the distribution of the train-set and test-set is a well-known issue in the domain of machine learning, and it is referred to as \textit{covariate shift}~\cite{covariate_shift}. 
\par
As stated above, typically, we do not know anything about the distribution of the attacked-set. \massimo{However, once} we crack the first password, we can start to observe and model the attacked distribution. Every new successful guess provides valuable information that we can leverage to improve the quality of the attack, \ie to reduce the \textit{covariate shift}. This iterative procedure recalls a Bayesian-like approach since there is continuous feedback between observations and the probability distribution. 
\par
For fully data-driven approaches, a naive solution to incorporate the acquired information from successful guesses is to fine-tune the model to change the learned password distribution. However, \textit{prescribed probabilistic models} such as FLA directly estimate the password distribution using a parametric function:
\begin{equation}
p(\x) = p(\x; \theta),
\end{equation}
where $\theta$ is the set of weights of a neural network. In this case, the only possibility of modifying the distribution $p(\x)$ in a meaningful way is to act on $\theta$ by harnessing the learning process. However, this is not an easy/attractive solution mainly because the new guessed passwords are potentially inadequate representatives\footnote{A very few guessed passwords against a dataset of millions of unknown passwords.} and will not force the model to generalize over the new information. Additionally, the computational cost of fine-tuning the network is considerable, and the final results cannot be guaranteed due to the sensitivity of the learning process.
\par
Similar to FLA, our generative model also exploits a neural network as an estimator. However, its modeled distribution \massimo{is} a joint probability distribution, shown in Eq.~\ref{equation:joint_probability}:
\begin{equation}
p(\x) = p(\x, \z) = p(\x \mid z; \theta) p(\z),
\label{equation:joint_probability}
\end{equation}
where $p(\z)$ is referred to as the latent distribution.
\par
As introduced in Section~\ref{section:gan}, when $p(\z) = \dot{p}(\z)$ (\ie prior latent distribution), $p(\x \mid z;\ \theta) p(\z)$ acts as a good approximation of the target data distribution (\ie the distribution followed by the train-set). Nevertheless, $p(\z)$ can be arbitrarily chosen and used to indirectly change the probability distribution modeled by the generator. The RHS of the Eq.~\ref{equation:joint_probability} clearly shows that $\theta$ is not the only free parameter affecting the distribution of the final passwords. Indeed, $p(\z)$ is completely independent of the generator, and so it can be modified arbitrarily without acting on the parameters of the neural network.
\par
This possibility, along with the passwords locality of the latent space, allows us to correctly and efficiently generalize over the new guessed passwords, leading the pre-trained network to model a password distribution closer to the guessed ones. It is noteworthy that this capability of generalizing over the new points is achieved via the weak locality and not from the neural network itself. \textbf{The intuition here is that when we change $p(\z)$ to assign more density to a specific guessed password $x$, we are also increasing the probability of its neighboring passwords that, due to the weak locality property, share similar characteristics.} This, in turn, makes it possible to \massimo{highlight} the general features of the guessed passwords (\eg structure, length, character set, \etc).
\begin{figure*}[!t]
\resizebox{1\textwidth}{!}{%
	\centering
	\subfigure[Actual attacked set]{
		\centering
		\includegraphics[trim = 35mm 35mm 35mm 35mm, clip, width=.16\linewidth]{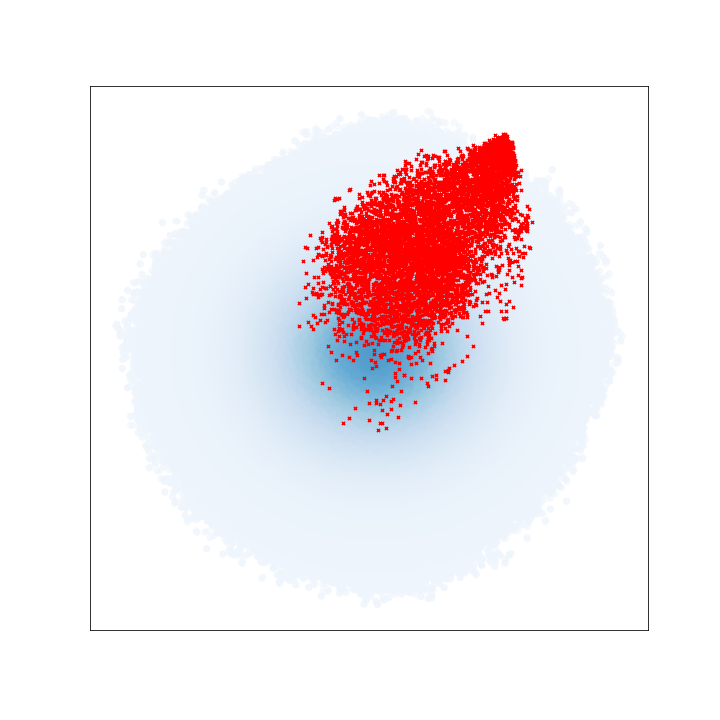}
	}\hspace{.15em}
	\subfigure[$10^4$ generation]{
		\centering
		\includegraphics[trim = 35mm 35mm 35mm 35mm, clip, width=.16\linewidth]{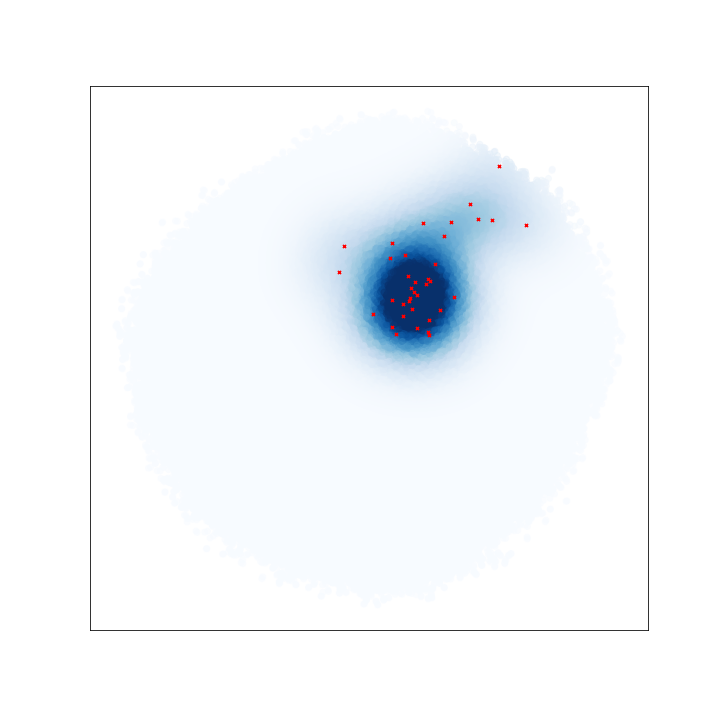}
	}\hspace{.15em}
	\subfigure[$10^5$ generation]{
		\centering
		\includegraphics[trim = 35mm 35mm 35mm 35mm, clip, width=.16\linewidth]{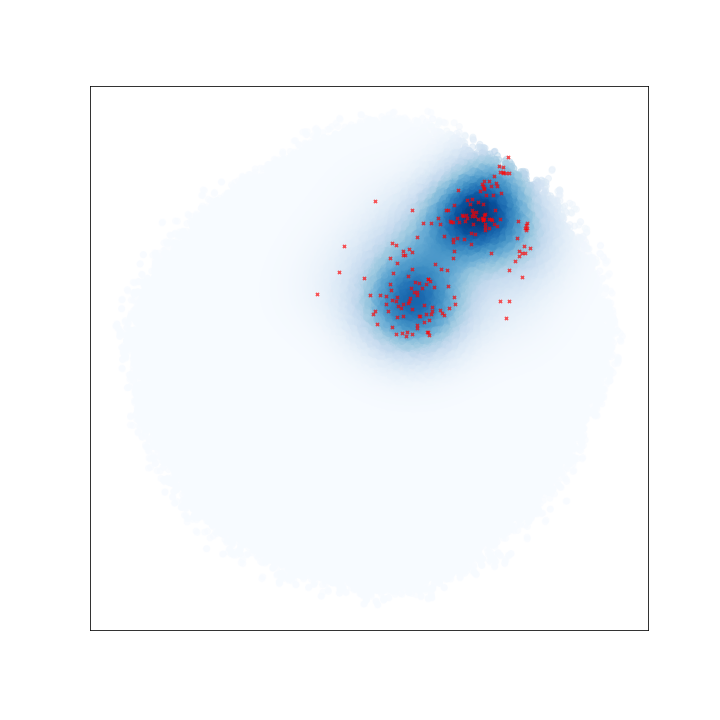}
	}\hspace{.15em}
	\subfigure[$10^6$ generation]{
		\centering
		\includegraphics[trim = 35mm 35mm 35mm 35mm, clip, width=.16\linewidth]{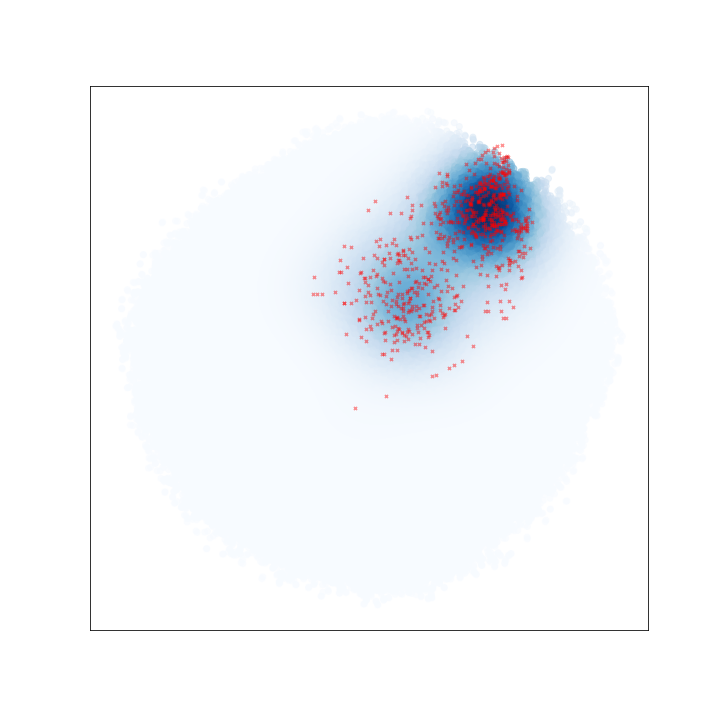}
	}\hspace{.15em}
	\subfigure[$10^7$ generation]{
		\centering
		\includegraphics[trim = 35mm 35mm 35mm 35mm, clip, width=.16\linewidth]{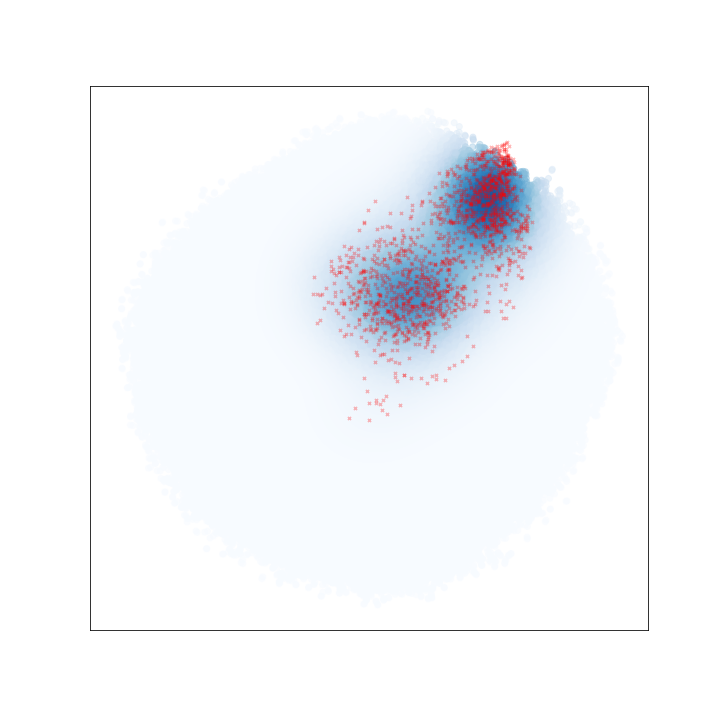}
	}
}
	\caption{2D visualization of: (a) the entire \textit{hotmail} dataset (red-part) mapped on the latent space learned from \massimo{the} RockYou train-set and (b-e) the latent space in four progressive attack steps for DPG on the \textit{hotmail} test-set. The red markers portray the guessed passwords at each step (\ie the $Z_i$), whereas the color intensity of the blue regions depicts the probability assigned from the used latent distribution (\ie mixture of Gaussians) to the latent space.}
	\label{figure:dynamic_attack_visulization}

\end{figure*}

\begin{figure*}[!htbp]
	\centering
	\subfigure{
		\centering
		\includegraphics[trim = 0mm 98mm 0mm 0mm, clip, width=.35\linewidth]{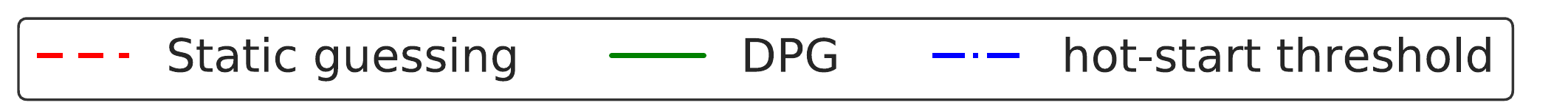}
	}\vspace{-.3em}
	\addtocounter{subfigure}{-1}
	
	\subfigure[\textit{myspace}]{
		\centering
		\includegraphics[trim = 15mm 0mm 2mm 15mm, clip, width=.25\linewidth]{./images/dynamic/myspace_8_dynamic_static.pdf}
	}
	\subfigure[\textit{hotmail}]{
		\centering
		\includegraphics[trim = 11mm 0mm 2mm 15mm, clip, width=.25\linewidth]{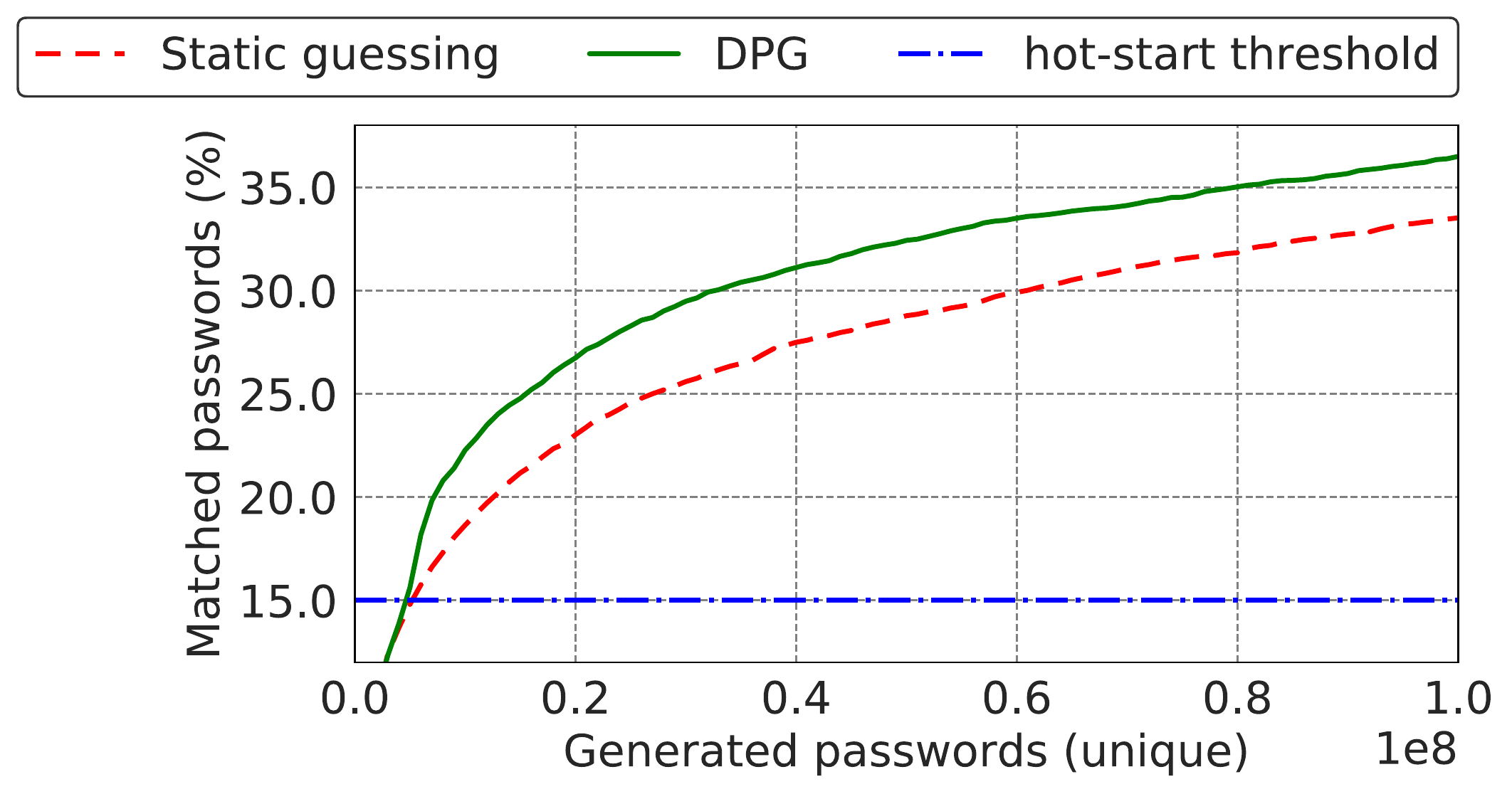}
	}
	\subfigure[\textit{phpbb}]{
		\centering
		\includegraphics[trim = 15mm 0mm 2mm 15mm, clip,
		width=.25\linewidth]{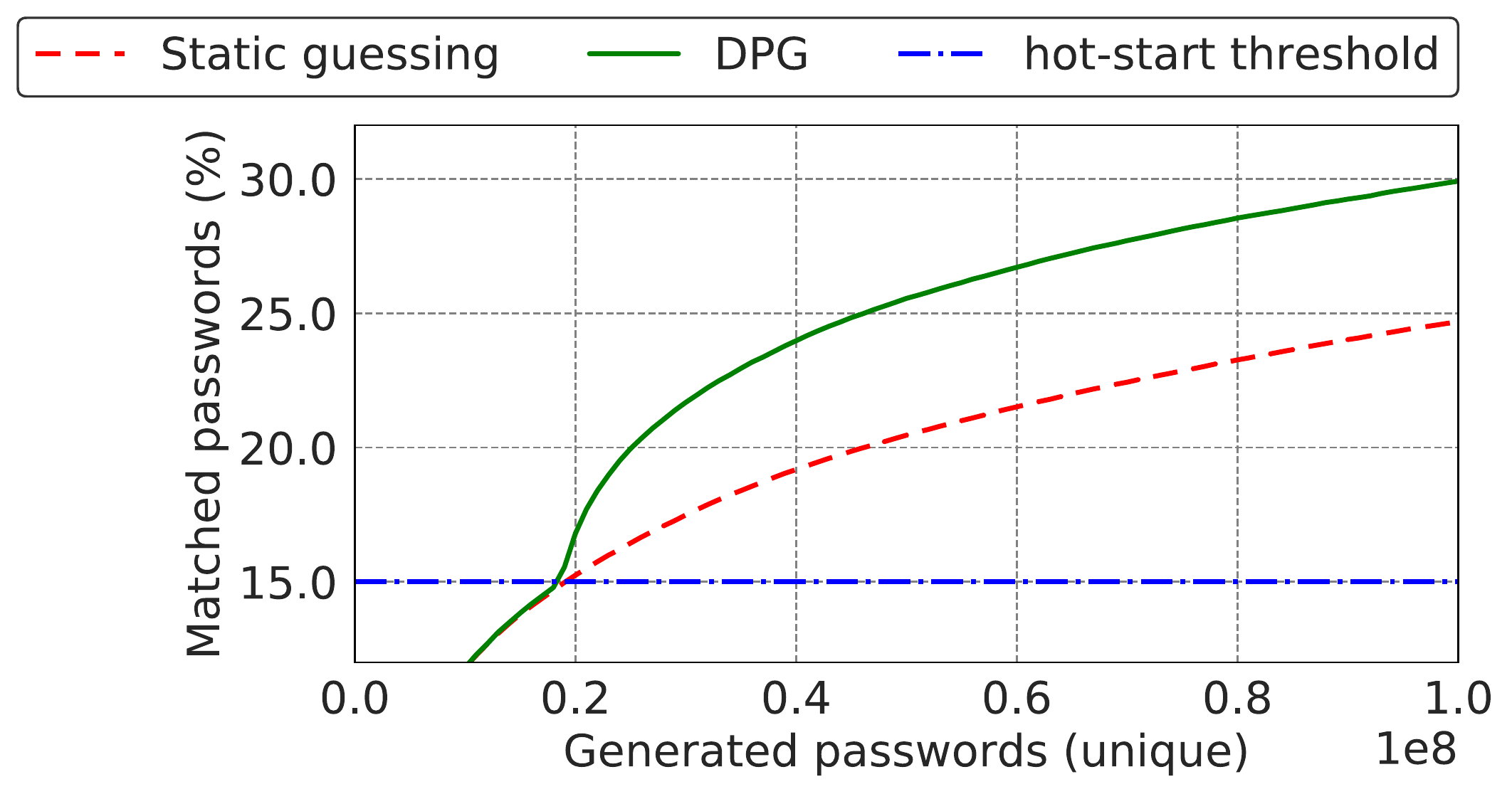}
	}
	\caption[]{The performance gain obtained by DPG (with $\alpha = 0.15$) with respect to static attack for three different test-sets}
	\label{figure:hot_start}
	\vspace{-1em}
\end{figure*}
\par
Thus, by controlling the latent distribution, we can increase the probabilities of the zones potentially covered by the passwords coming from the target distribution. We call this technique \textbf{Dynamic Password Guessing (DPG)}. In the case of homogeneous distribution (\eg\textit{myspace}), we can narrow down the solution space around the dense zones, and avoid exploring the entire latent-space. On the other hand, for passwords sets sampled from distributions far from the one modeled by the generator, we can focus on zones of the latent space, which, otherwise, would have been poorly explored. In both cases, we can reduce the \textit{covariate shift} and improve the performance of the password guessing attack.
\par
\textbf{In a broad sense, DPG can potentially adapt to very peculiar password distributions; distributions induced from the contexts where no suitable train-sets can be collected. \Eg passwords created under an unmatched composition policy or rare/unobserved users' habits. As long as the generator has a non-zero probability of generating such rare passwords, the feedback given from the correct guesses can consistently be used to reweigh the latent distribution and mimic the unknown target password distribution.} We will validate this claim in the next section.
\begin{figure*}[!b]
	\centering
	
	\subfigure{
		\centering
		\includegraphics[trim = 25mm 92mm 20mm 0mm, clip, width=.8\linewidth]{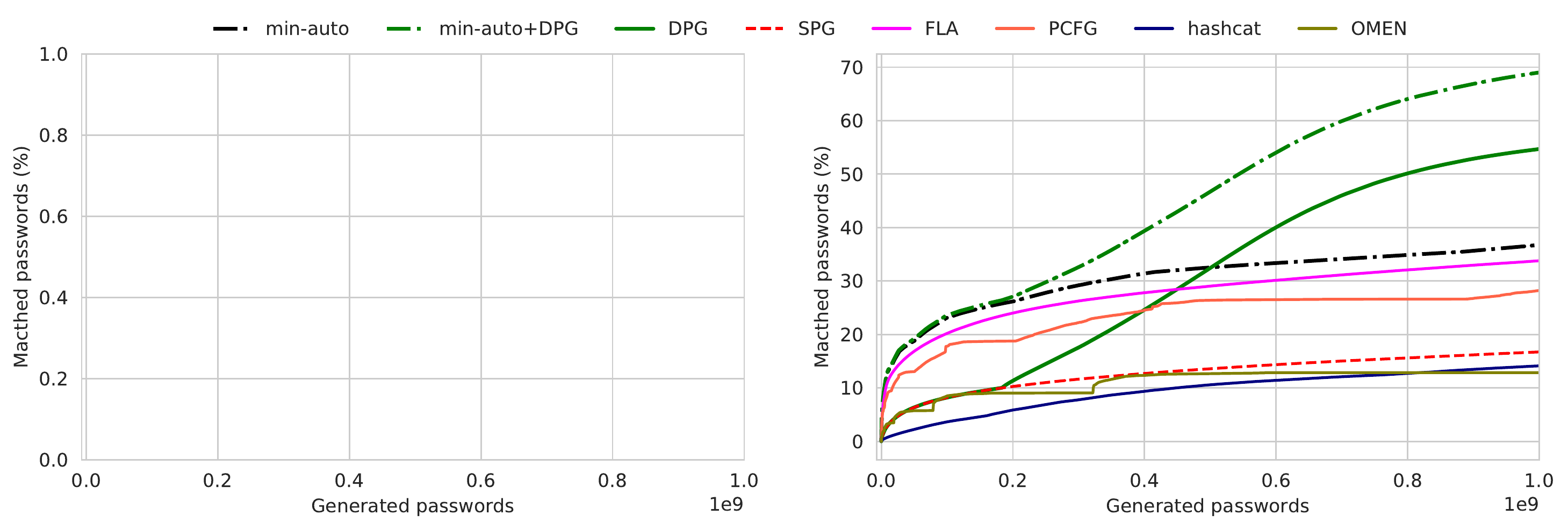}
	}\vspace{-.5em}
	\addtocounter{subfigure}{-1}
	
	\subfigure[\textit{LinkedIn}]{
		\includegraphics[trim = 0mm 0mm 0mm 0mm, clip, width=.28\linewidth]{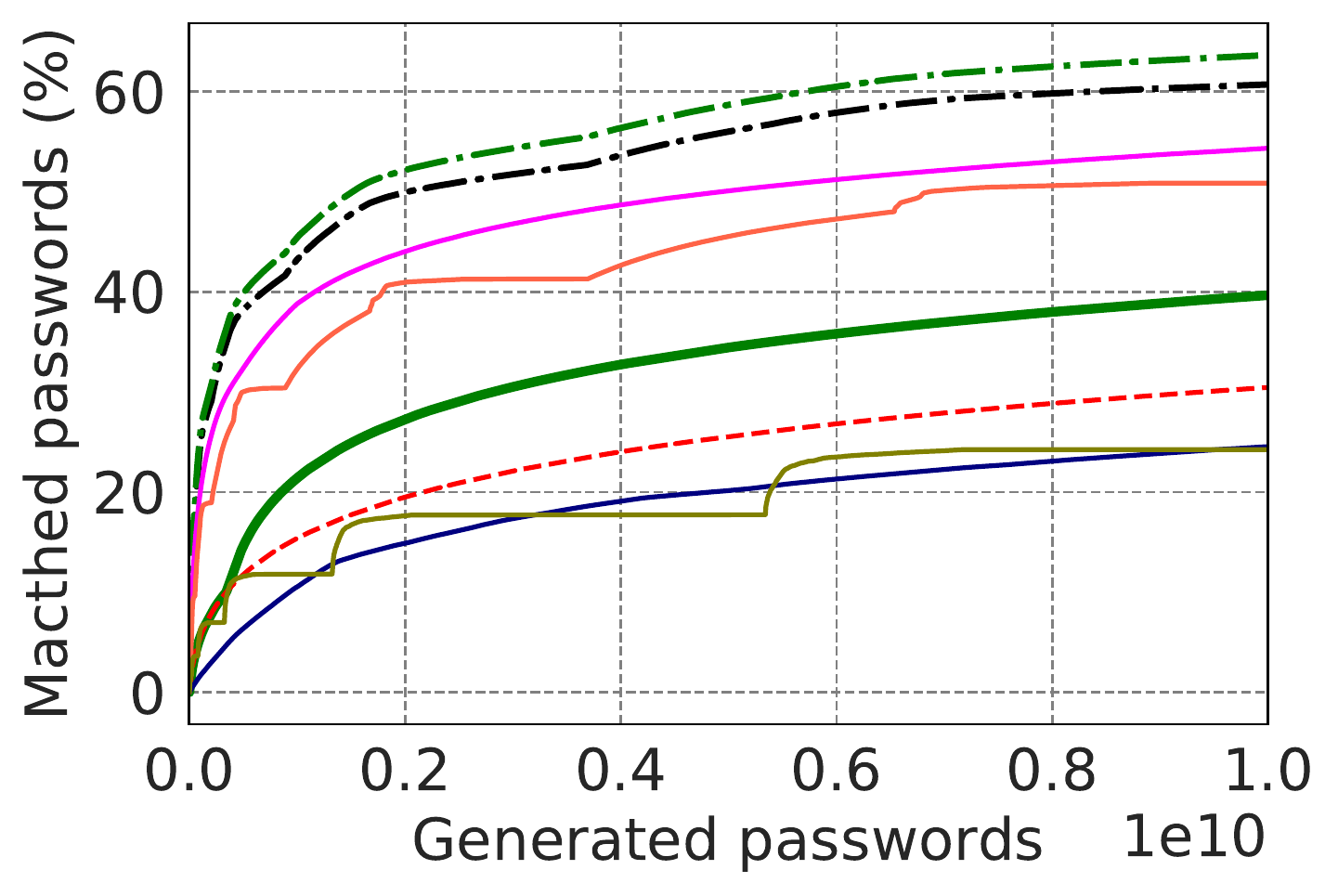}
	}\subfigure[\textit{Youku}]{
		\includegraphics[trim = 0mm 0mm 0mm 0mm, clip, width=.28\linewidth]{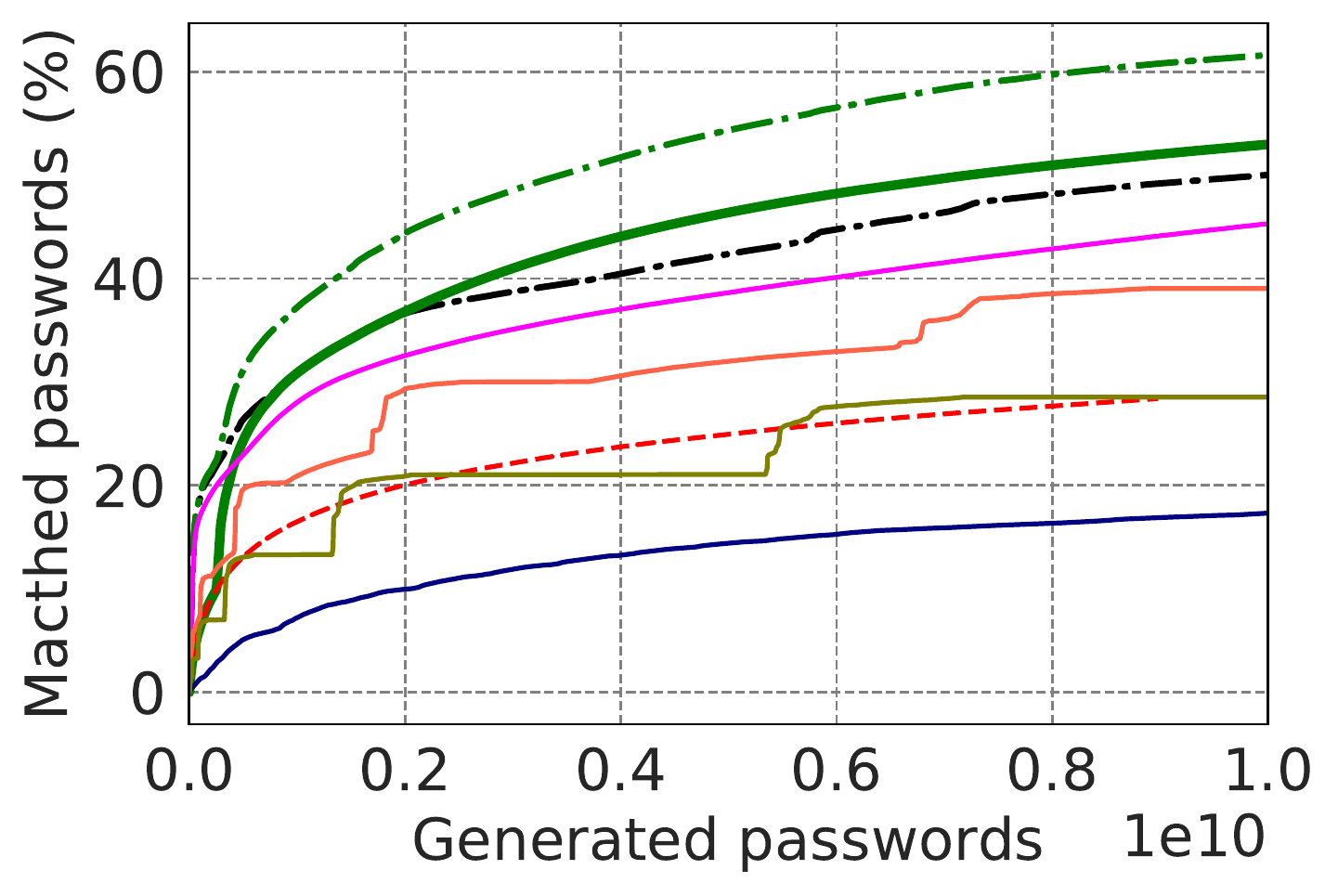}
	}\subfigure[\textit{Zomato}]{
		\includegraphics[trim = 0mm 0mm 0mm 0mm, clip, width=.28\linewidth]{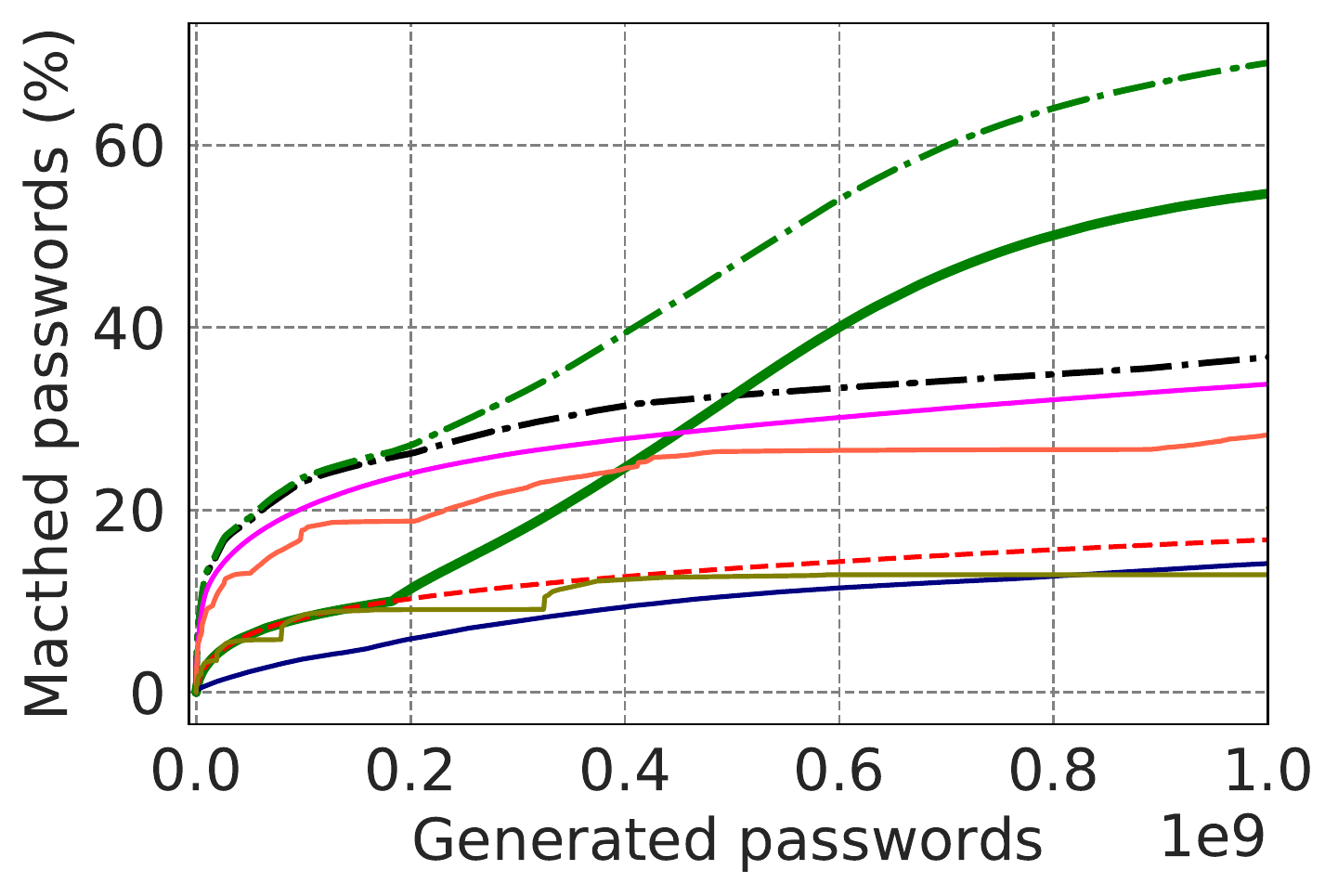}
	}
	\caption{Performance of various password models on three password leaks. For DPG, we used $\sigma=0.35$ and $\alpha=10\%$.}
	\label{fig:linkzom}
	\vspace{-1em}
\end{figure*}
\subsubsection{Practical implementation}
\label{section:dynamic_attack_in_practice}
\par
In this section, we cover DPG from a practical viewpoint. Algorithm~\ref{algorithm:dynamic_attack} briefly describes DPG.
\begin{algorithm}[b]
	\caption{Dynamic Password Guessing (DPG)}
	\label{algorithm:dynamic_attack}
	\scriptsize
	\begin{algorithmic}[1]
		\renewcommand{\algorithmicrequire}{\textbf{Input:}}
		\renewcommand{\algorithmicensure}{\textbf{Output:}}
		\REQUIRE Set: $\mathcal{O}$, Int: $\alpha$
		\STATE $i=0$
		\STATE $p_{\text{latent}} = \dot{p}(\x)$
		\STATE $Z = \{\}$
		\FOR {\!\!\textbf{each} $z \sim p_{\text{latent}}$}
		\STATE$x = G(z)$
		\IF{$x \in \mathcal{O}$}
		\STATE $i++$
		\STATE $Z_i = Z = Z \cup \{z\}$
		\IF{$i \geq \alpha$}
		\STATE $p_{\text{latent}} = \text{makeLatentDistribution}(Z_i)$
		\ENDIF
		\ENDIF
		\ENDFOR
	\end{algorithmic}
\end{algorithm}
\vspace{-1em}
\par
In Algorithm~\ref{algorithm:dynamic_attack}, $\mathcal{O}$ represents the target set of passwords, $Z$ is the collection of all the passwords guessed by the generator, and $\alpha$ is defined as \massimo{the} \textbf{hot-start} parameter of the attack, an element that we describe later in this section. The variable $p_{\text{latent}}$ in the pseudo-code, represents the latent distribution from which we sample latent points. 
The procedure \textit{makeLatentDistribution} returns the latent distribution induced from the group of guessed passwords $Z_i$ at step $i$. Leveraging the maximum-likelihood framework, we choose such a distribution to maximize the probability of the set of observed passwords $X_i~\myeq~\{G(z) \mid z \in Z_i\}$. 
This is accomplished by considering a latent distribution $p(\z\mid Z_i)$ conditioned to the set of passwords guessed at each step $i$. The final password distribution represented by the generator during DPG is reported in Eq.~\ref{equation:condition}.
\begin{equation}
p(\x) = p(\x \mid z; \theta) p(\z \mid Z_i).
\label{equation:condition}
\end{equation}
\par
As a natural extension of the proximity password generation harnessed in Section~\ref{section:password_template_inversion}, we choose to represent $p(\z~|~Z_i)$ as a finite mixture of isotropic Gaussians. In particular, the mixture is composed of $n$ Gaussians, where:~(1)~$n$ is the number of the latent points in $Z_i$; and (2)~for each $z_j \in Z_i$, a Gaussian is defined as $\gaussian(z_j, \sigma \mathbf{I})$ with center as $z_j$ and a fixed standard deviation $\sigma$.
\par
When the probability of a password, \ie $x_j = G(z_j)$, is known, we weight the importance of the $j^{th}$ distribution as $P(x_j)$; otherwise a uniform distribution among the Gaussians is assumed. {In the reported experiments, we always used uniform weighting. 
Equation~\ref{equation:pdf_of_latent_space} defines the probability density function of the latent space.
\begin{equation}
\label{equation:pdf_of_latent_space}
p(\z \mid Z_i)= \sum_{j=0}^{n} P(G(z_j))\cdot \gaussian(\z \mid z_j, \sigma \mathbf{I}).
\end{equation}

Every new guessed password $x$ introduces a new Gaussian centered at $z$ to the mixture. \massimo{Consequently, every new guessed password contributes to changes in the latent distribution $p(\z~|~Z_i)$ by moving the density of the distribution in the zone of the latent space where it lies. \figurename{~\ref{figure:dynamic_attack_visulization}} visualizes this phenomenon.}

\par
In the context of DPG, the GAN generator performs slightly better than CWEA. For this reason, all the experiments reported in this section are obtained with our GAN generator trained on the RockYou train-set.
\figurename{~\ref{figure:hot_start}} depicts the performance comparison between a static attack (\eg PassGAN) and DPG over the three passwords sets. Adaptively changing the latent distribution allows us to boost the number of guessed passwords per unit of time. 
Importantly, this improvement comes without any additional information or assumption over the attacked passwords set. In \massimo{addition, the computational overhead} due to the new sampling technique is negligible. The steep improvement in the performance obtained with DPG supports our view that reducing the \textit{covariate shift} is a sound~strategy.

\par 
The sudden growth in the guessed passwords in DPG (shown in~\figurename{~\ref{figure:hot_start}}) is due to the hot-start or $\alpha$ parameter;
in DPG, we use the prior latent distribution until a predetermined number ($\alpha$) of passwords has been guessed. After that, we start to use the conditional latent distribution $p(\mathbf{z}~\mid~Z_i)$. 
The reason is that if DPG starts with the very first guessed password, then the latent distribution can be stuck in a small area of the latent space. However, launching DPG after guessing a sufficient number of passwords (\ie after finding a set of uncorrelated latent points in the latent space) gives us the possibility to match a heterogeneous set of passwords, which correctly localize the dense zones of the latent space where the attacked passwords are likely to lie.

\par
\begin{figure*}[!b]
	\centering
	\subfigure[Probability according to train-set]{
		\centering
		\includegraphics[width=.260\linewidth]{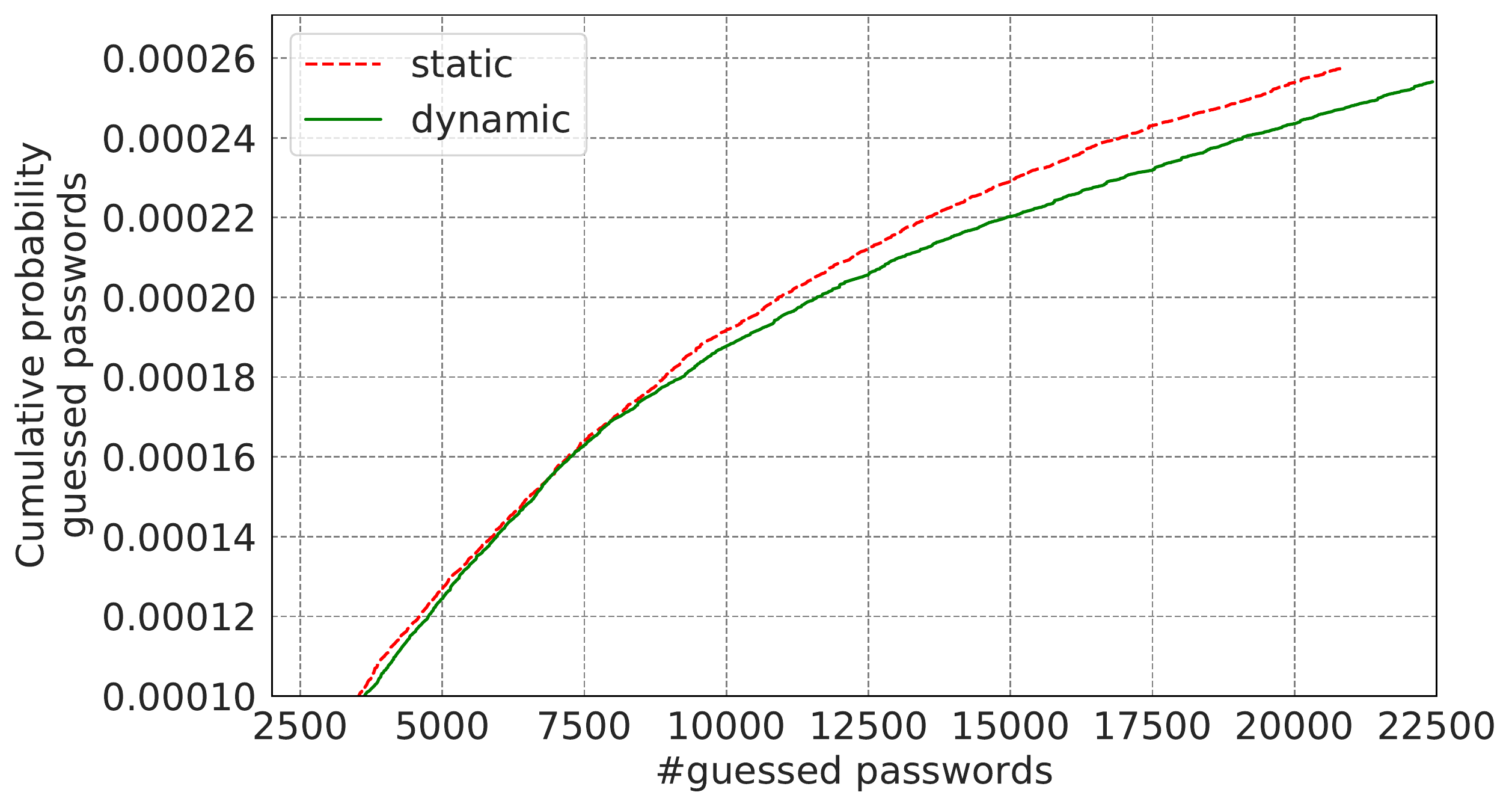}
	}
	\subfigure[Probability according to attackes-set]{
		\centering
		\includegraphics[width=.260\linewidth]{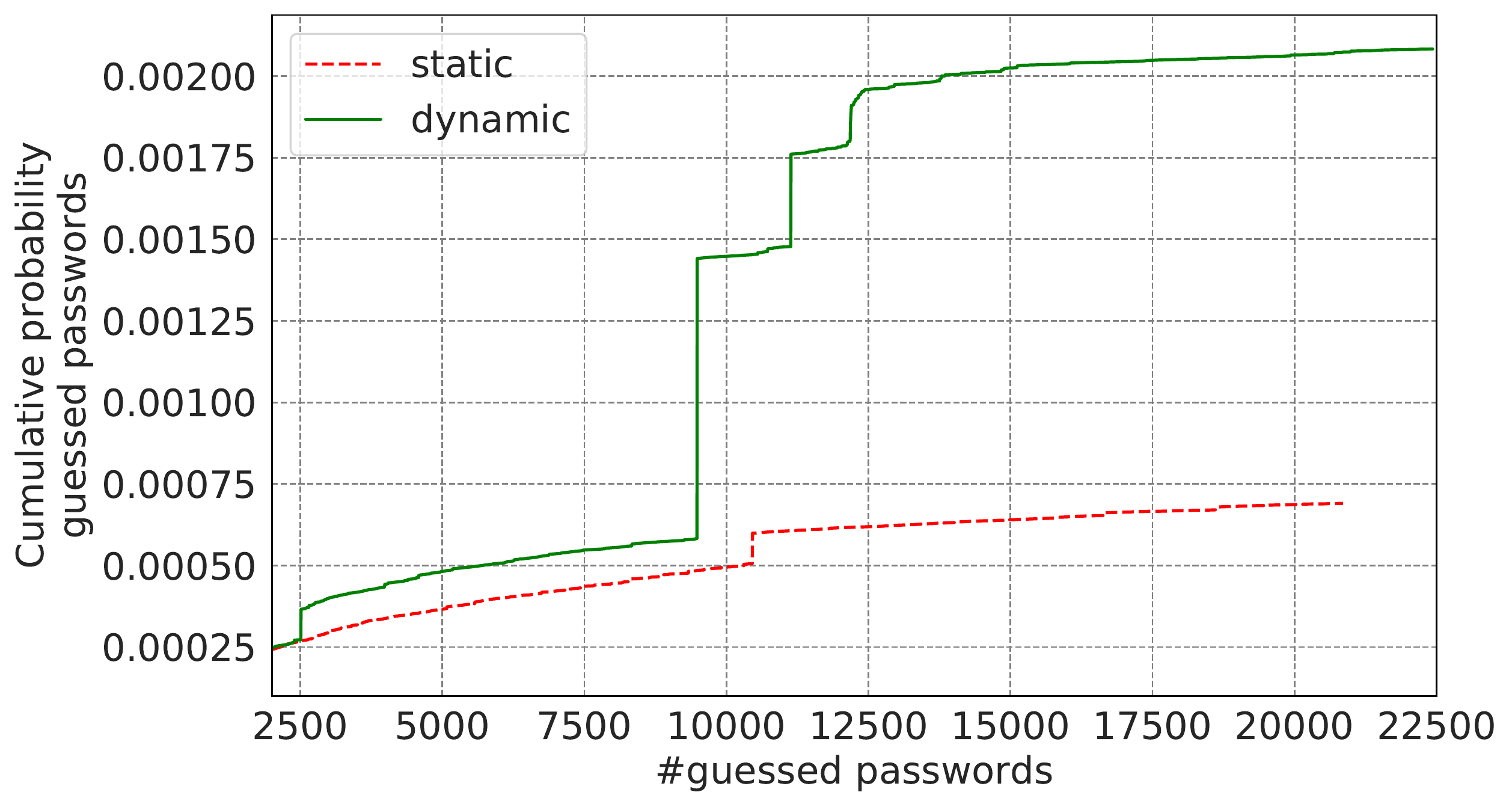}
	}
	\subfigure[Guess-number according to train-set]{
		\centering
		\includegraphics[width=.250\linewidth]{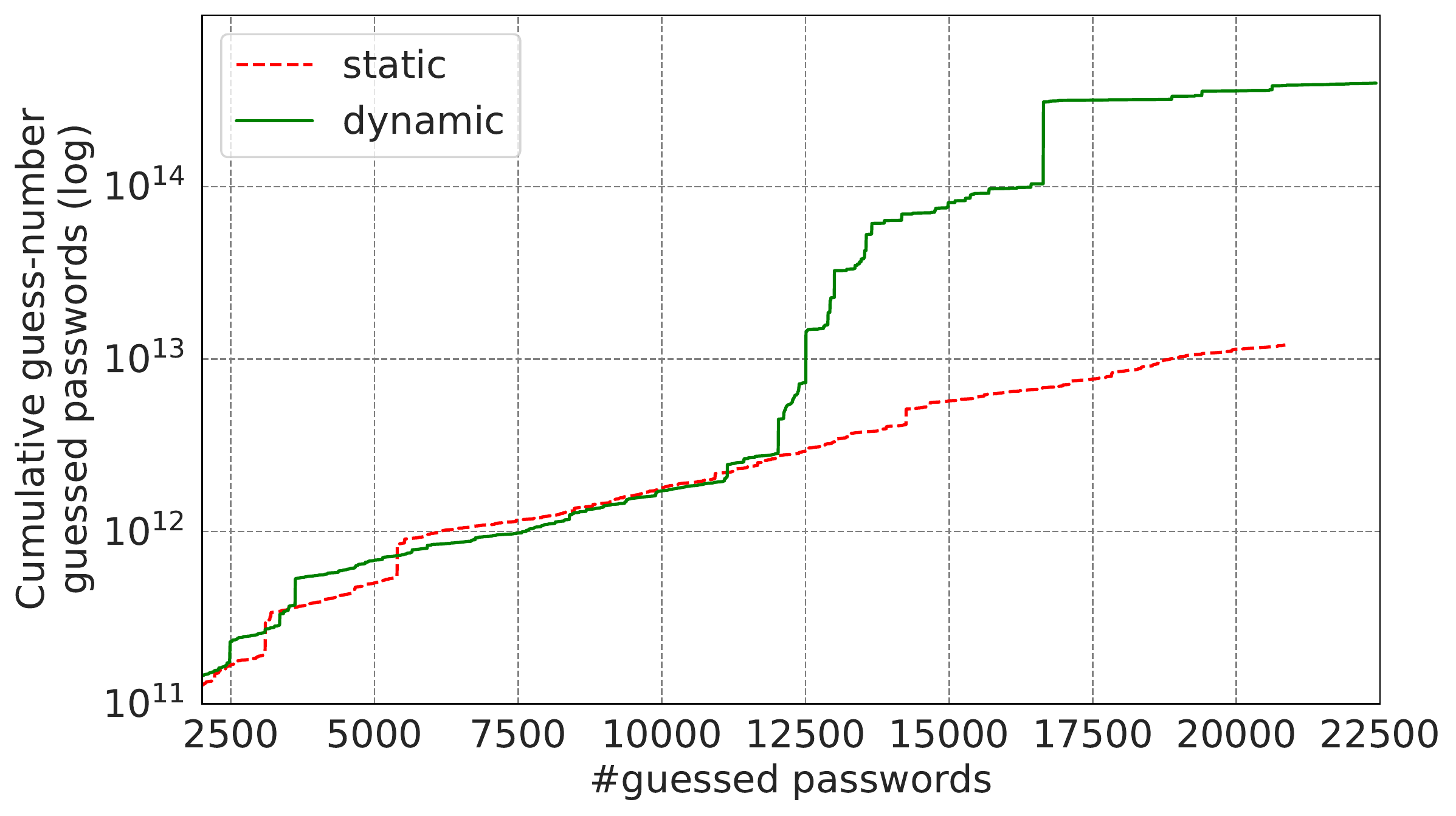}
	}
	\caption[]{Cumulative statistic for a guessing attack over \textit{phpbb}. The figures report the password guessed in the first $10^9$ guesses for both static and dynamic.}
	\label{figure:dynamic_phpbb_pg}
	\vspace{-1em}
\end{figure*}
The final hyper-parameter of our attack is the standard deviation ($\sigma$) assigned to every Gaussian in the mixture. Under the Kernel Density Estimation (KDE) perspective, $\sigma$ represents the \textit{bandwidth} of our Gaussian kernels. In the guessing scenario, instead, this value defines how far we want to sample from the clusters of observed passwords. A larger value of $\sigma$ allows us to be less biased and explore a wider zone around the guessed passwords; whereas a smaller value enables a more focused inspection of the latter. Appendix \ref{sec:dpg_h} better explicates the effect of $\sigma$ and $\alpha$ on DPG.

\keepme{
In \figurename{~\ref{fig:linkzom}}, we report a direct comparison of the proposed DPG against \sota password models for three password leaks. For the comparison, we used the same tools and configurations described in Section~\ref{sec:eval_cpg}.\footnote{For the \textit{min-auto}, we do not use CMU-PGS~\cite{cmupgs} directly given their limits on the number of queries allowed and the cardinality of the tested sets.} In the figure, DPG refers to the dynamic guessing attack, whereas SPG to the static one. \textit{min-auto} is obtained by combining the guesses of FLA, Hashcat, OMEN, and PCFG. \textit{min-auto+DPG} is then obtained by adding DPG to the \textit{min-auto} ensemble. \figurename{~\ref{fig:linkzom}~(a)} reports the results for the \textit{LinkedIn} leak. Here, the dynamic adaptation allows us to guess up to $10\%$ more passwords then the static approach. However, it cannot directly match the performance of FLA and PCFG in this general case. Nevertheless, our models behave better than mangling rules and the Markov model. Given the different nature of the dynamic guessing strategy, combining DPG with \textit{min-auto} permits us to guess more passwords. We will better motivate this phenomenon in the next section.
}
\par
\begin{table*}[t]
	\centering
	\caption{Example of peculiar passwords guessed via DPG for four password leaks. The required numbers of guesses (\ie \textbf{G}) are reported for both FLA and our DPG. These passwords have been obtained by ordering all the guessed passwords of the DPG attacks in decreasing order based on the guess-number assigned from FLA . The table reports the first $15$ entries of the list for each leak.}
	\label{table:peculiar}
	\resizebox{1\textwidth}{!}{%
		\begin{tabular}{cc|ccccccccccccccc}
			\toprule
			\textit{LinkedIn} & Guess & o2linkedln & w2linkedln & ydlinkedln & linked6in6 & j*linkedln & linked!in. & wslinked1n & linkedgcin & linked6in2 & lslinkedln & wtlinkedln & 9auiirji & g2linkedln & cslinkedln & ymlinkedln \\
			&FLA \textbf{G} & $8.2\cdot 10^{15}$ & $6.3\cdot 10^{15}$ & $3.6\cdot 10^{15}$ & $3.6\cdot 10^{15}$ & $3.0\cdot 10^{15}$ & $2.8\cdot 10^{15}$ & $2.6\cdot 10^{15}$ & $2.5\cdot 10^{15}$ & $1.4\cdot 10^{15}$ & $1.4\cdot 10^{15}$ & $1.3\cdot 10^{15}$ & $1.2\cdot 10^{15}$ & $1.2\cdot 10^{15}$ & $1.1\cdot 10^{15}$ & $1.0\cdot 10^{15}$\\
			&DPG \textbf{G} & $3.4\cdot 10^{9}$ & $3.1\cdot 10^{9}$ & $3.6\cdot 10^{9}$ & $4.3\cdot 10^{9}$ & $4.3\cdot 10^{9}$ & $4.8\cdot 10^{9}$ & $4.4\cdot 10^{9}$ & $2.1\cdot 10^{9}$ & $5.6\cdot 10^{9}$ & $4.5\cdot 10^{9}$ & $4.5\cdot 10^{9}$ & $5.5\cdot 10^{9}$ & $3.4\cdot 10^{9}$ & $4.4\cdot 10^{9}$ & $5.2\cdot 10^{9}$ \\
			\midrule
			\textit{Youku} & Guess & guoxuange2 & xuhaidong7 & caoxia521. & woailc521. & woyijiu521 & woaicyhx0 & xuhaidong1 & woaifiy520 & yishwng521 & woshiqujie & yixuan520. & slllong520 & woaifuyao & woshiqutao & liu19981.6  \\
			&FLA \textbf{G} & $2.5\cdot 10^{15}$ & $1.7\cdot 10^{15}$ & $1.3\cdot 10^{15}$ & $9.6\cdot 10^{14}$ & $7.3\cdot 10^{14}$ & $6.5\cdot 10^{14}$ & $6.4\cdot 10^{14}$ & $6.4\cdot 10^{14}$ & $5.3\cdot 10^{14}$ & $5.1\cdot 10^{14}$ & $5.0\cdot 10^{14}$ & $5.0\cdot 10^{14}$ & $4.9\cdot 10^{14}$ & $4.7\cdot 10^{14}$ & $4.6\cdot 10^{14}$  \\
			&DPG \textbf{G} & $3.2\cdot 10^{9}$ & $3.9\cdot 10^{9}$ & $3.5\cdot 10^{9}$ & $3.7\cdot 10^{9}$ & $3.5\cdot 10^{9}$ & $3.3\cdot 10^{9}$ & $3.9\cdot 10^{9}$ & $3.8\cdot 10^{9}$ & $3.7\cdot 10^{9}$ & $3.0\cdot 10^{9}$ & $3.8\cdot 10^{9}$ & $3.2\cdot 10^{9}$ & $1.4\cdot 10^{9}$ & $2.4\cdot 10^{9}$ & $1.3\cdot 10^{9}$ \\
			\midrule
			\textit{Zomato} & Guess & z0mato2016 & z0mato2015 & zomato9a00 & 2defd0 & zomat\_997 & 3aee0f & zomato\_496 & zomato\_443 & zomato.921 & zomato\_591 & zomato\_543 & 0def0a & zomato\_441 & zomato\_948 & zomato\_376 \\
			&FLA \textbf{G} & $1.9\cdot 10^{14}$ & $1.5\cdot 10^{14}$ & $1.2\cdot 10^{14}$ & $7.3\cdot 10^{13}$ & $4.0\cdot 10^{13}$ & $3.8\cdot 10^{13}$ & $3.5\cdot 10^{13}$ & $3.4\cdot 10^{13}$ & $3.2\cdot 10^{13}$ & $3.1\cdot 10^{13}$ & $3.1\cdot 10^{13}$ & $3.0\cdot 10^{13}$ & $2.9\cdot 10^{13}$ & $2.9\cdot 10^{13}$ & $2.8\cdot 10^{13}$ \\
			&DPG \textbf{G} & $4.5\cdot 10^{8}$ & $7.7\cdot 10^{8}$ & $7.8\cdot 10^{8}$ & $5.1\cdot 10^{8}$ & $1.0\cdot 10^{9}$ & $8.1\cdot 10^{8}$ & $8.0\cdot 10^{8}$ & $1.1\cdot 10^{9}$ & $1.1\cdot 10^{9}$ & $1.1\cdot 10^{9}$ & $1.0\cdot 10^{9}$ & $4.5\cdot 10^{8}$ & $1.1\cdot 10^{9}$ & $1.1\cdot 10^{9}$ & $1.2\cdot 10^{9}$ \\
			\midrule
			\textit{phpbb} & Guess &phpbb3.14 & phpbb0472 & phpbb4s2 & phpbb7825 & phpbbid12 & phpbb8424 & phpbb3546 & phpbb4291 & phpbb8686 & phpbb9801 & phpbb1902 & phpbb5682 & sksdbles & phpbb1298 & phpbb2625\\  
			&FLA \textbf{G} & $2.1\cdot 10^{14}$ & $2.1\cdot 10^{13}$ & $2.0\cdot 10^{13}$ & $1.3\cdot 10^{13}$ & $1.0\cdot 10^{13}$ & $9.9\cdot 10^{12}$ & $8.0\cdot 10^{12}$ & $7.2\cdot 10^{12}$ & $5.5\cdot 10^{12}$ & $5.4\cdot 10^{12}$ & $4.5\cdot 10^{12}$ & $4.5\cdot 10^{12}$ & $3.8\cdot 10^{12}$ & $3.8\cdot 10^{12}$ & $3.2\cdot 10^{12}$ \\
			&DPG \textbf{G} & $2.4\cdot 10^{8}$ & $6.5\cdot 10^{8}$ & $4.8\cdot 10^{8}$ & $4.2\cdot 10^{8}$ & $1.2\cdot 10^{8}$ & $1.1\cdot 10^{8}$ & $1.3\cdot 10^{8}$ & $1.0\cdot 10^{8}$ & $1.4\cdot 10^{8}$ & $2.0\cdot 10^{8}$ & $2.3\cdot 10^{8}$ & $1.7\cdot 10^{8}$ & $1.2\cdot 10^{8}$ & $1.8\cdot 10^{8}$ & $1.1\cdot 10^{8}$ \\
			\bottomrule
		\end{tabular}
	}
\end{table*}
\keepme{
Consistently better results are observed as soon as we consider leaks that exhibit peculiar biases in their password distributions. \figurename{~\ref{fig:linkzom}~(b)} reports the results for the leak \textit{Youku}~\cite{hashes_leak, hacked_read} - a Chinese video hosting service. In this case, the inherent distribution shifts induced by a different class of users causes a substantial covariate shift phenomenon. Here, the dynamic adaptation allows us to guess more passwords than the other tools; DPG improves guess after guess, each time evolving and eventually surpassing the \textit{min-auto} configuration obtained by combining all other models.}
\par
\keepme{
Even more interesting results can be observed when we consider leaks that introduce heavier biases. \figurename{~\ref{fig:linkzom}~(c)} reports the results for the \textit{Zomato}~\cite{zomato_leak, et_zomato} leak. This leak is an extreme case since $\sim40\%$ of its content includes random tokens of six alphanumeric characters. 
That creates a sharply segmented bimodal distribution that can be detected and efficiently captured by DPG. In this instance, the dynamic adaptation of the latent space allows us to guess up to $\sim5$ times more passwords than the static attack (\ie SPG), allowing our model to match more than $50\%$ of the set in less than $10^9$ iterations. On the other hand, static approaches, including \textit{min-auto}, cannot match the performance of DPG in this extreme case. Of note, adding DPG to the ensemble of \textit{min-auto} (\ie \textit{min-auto+DPG}) allows us to guess $\sim70\%$ of the set}.
\par
\keepme{
The last two examples highlight the ability of DPG to adapt to the target password distribution. However, the result of the \textit{LinkedIn} leak tells us that the dynamic attack cannot directly match the performance of \sota solutions in case there is no evident covariate shift. 
In the next section, we will show that the DPG algorithm is indeed useful also in such cases, as it soundly permits to guess peculiar passwords of the attacked distribution that would be otherwise ignored.}
\par

\subsubsection{The impact of the dynamic adaptation}
\keepme{
	In this section, we clarify the effect of the dynamic latent adaptation over the password distribution originally modeled from the deep generative model. To this end, we compare the probability of the guessed passwords according to different password distributions, namely, (1)~the distribution of the train-set and (2)~the distribution of the attacked-set of passwords. To soundly represent and generalize such probability distributions, we rely on FLA~\cite{fla} as an explicit password mass estimator. We train two instances of FLA on the two passwords sets and use the trained models to infer probabilities over the password guessed during the dynamic and static attacks.}
\par
\keepme{
	\figurename{~\ref{figure:dynamic_phpbb_pg}} summarizes our measurements for the \textit{phpbb} password leak (\ie the attacked distribution). Here, the cumulative probability of the guessed password is reported for both dynamic and static attacks. In particular, \figurename{~\ref{figure:dynamic_phpbb_pg}~(a)} describes the probabilities assigned from the probability distribution of the train-set (\ie the FLA instance trained on \textit{RockYou}), whereas \figurename{~\ref{figure:dynamic_phpbb_pg}~(b)} reports the same data points, but computed according to the probability distribution of the attacked-set (\ie the FLA instance trained on \textit{phpbb}).}
\par
\keepme{
	When we perform DPG, we expect the password distribution represented from the deep generative model to gradually diverge from the one learned at training time. \figurename{~\ref{figure:dynamic_phpbb_pg}~(a)} graphically describes this phenomenon; here, we note how the latent adaptation is causing the model to guess passwords that have a lower probability according to the train-set distribution. More interestingly, whereas the discrepancy between the modeled and the train distribution grows, the discrepancy sharply reduces for the attacked distribution. \figurename{~\ref{figure:dynamic_phpbb_pg}~(b)} explicates the convergence process towards the latter. Furthermore, this figure gives us a piece of more valuable information. It shows that the DPG guesses passwords that have high-probability according to the attacked distribution, \ie passwords associated with a higher number of users in the attacked service. Sudden jumps in the latter cumulative probability curve, indeed, can be attributed to the event of guessing such high-probability passwords. To note, once we guess a first high-probability password, we start sampling new guesses around it, guessing more high-probability passwords consequently and making those jumps even steeper.}}
\par
\keepme{
	Relying on the same example, more practical results can be appreciated when we consider the adversarial interpretation. \figurename{~\ref{figure:dynamic_phpbb_pg}~(c)} reports the cumulative guess-number graph for the static and dynamic attacks measured using the FLA instance trained on \textit{RockYou} (\ie the train-set of our model). The estimated cumulative guess-number of the dynamic attack is two magnitudes larger than that of the static attack. Considering FLA's accuracy~\cite{golla2018accuracy}, this result confirms that DPG can induce the generation of passwords that have low belief according to the train-set distribution. Moreover, this example shows how DPG can induce the earlier generation of passwords that would require multiple magnitude more guesses to be produced for equivalent state-of-the-art password guessers, such as FLA. In the reported example, we generate $10^9$ guesses, matching several passwords that would require up to $10^{14}$ iterations from FLA (and others; see Appendix \ref{section:supplementary_table_and_figures}). Table~\ref{table:peculiar} reports some of those.\\}

\keepme{
	We replicated the same analysis on different password leaks, observing the same general behavior.
	We reported high-guess-number passwords for those other sets as additional examples in Table~\ref{table:peculiar}. The listed guesses in the table give a clear intuition over the nature of such peculiar passwords. These are induced from unique biases of the attacked distribution. More evident examples are the passwords based on the name of web services that dominate the table. These are indeed the prime examples of peculiar passwords, as they univocally bound to the specific password distribution. More heterogeneous guesses can be observed in the row dedicated to the \textit{Youku} leak. Here, DPG captured passwords composed of peculiar dictionary entries that are not well represented in the train-set of the model (\ie \textit{RockYou}).\\
	Additionally, the guess-numbers reported in Table~\ref{table:peculiar} indicate that these are passwords that are considered secure by state-of-the-art tools, but that can be easily guessed through DPG.
	Indeed, our experiments show that \textbf{DPG allows us to guess passwords that are unique to the attacked password set. Such passwords, given their arbitrary distance from the general password distribution, can be soundly guessed only by leveraging additional sources of information over the attacked password space. DPG distills this necessary knowledge directly through an unsupervised interaction with the attacked set. This allows the guessing attack to automatically focus on unique modalities of the target password distribution that would otherwise be under-represented or ignored.}}

\section{Related Work}
\label{section:related_works}
Systematic studies on password guessing date back to 1979~\cite{morris1979password}, and probably, password guessing attacks have existed since the inception of the concept of passwords~\cite{bidgoli2006handbook}. Since a vast number of works have been proposed in this active area of research, we limit the discussion to the most relevant contributions and solutions that are highly related to our proposals.
\par
Dictionary-based attack and its extensions were among the first forms of elegant guessing techniques (as opposed to brute-forcing). Among dictionary attacks, the extension with mangling-rules~\cite{morris1979password} widely demonstrated its effectiveness on the trawling attack scenario~\cite{cracking_passwords_101}. Despite its simplicity, this attack approach persists nowadays in the form of highly tuned off-the-shelf software: John The Ripper~(JTR)~\cite{jtr} and HashCat~\cite{hashcat}. Due to their efficiency and easy customization, these tools are the primary weapons of professional security practitioners~\cite{ur2015measuring}.
Subsequently, probabilistic approaches naturally found their application in the password guessing domain. Narayanan~\etal~\cite{narayanan2005fast} apply a Markovian filter to reduce the searching space of a guessing attack drastically. D{\"u}rmuth~\etal~\cite{durmuth2015omen} extend that approach by introducing an improved version of the guesses enumeration algorithm in OMEN. 
Weir~\etal~\cite{pcfg} introduced Probabilistic Context-Free Grammars (PCFGs) in the password guessing domain. In particular, Weir~\etal~proposed a technique capable of inferring grammars from a set of observed passwords and use those to cast new password guesses. 
\par
Ciaramella~\etal~\cite{ciaramella2006neural} introduced neural networks for password guessing in their seminal work. In the same line of development, Melicher~\etal~\cite{fla} proposed FLA (Fast, Lean, and Accurate) that uses recurrent neural networks~\cite{graves2013generating, sutskever2011generating} to estimate the password distribution. This model follows the same estimation procedure of Markov models but relaxes the underlying $n$-markovian assumption. FLA can generate new guesses by performing an enumeration of the password space via a tree traversal algorithm. 
\par
Similarly to our conditional generation framework, different works have focused on creating a specific class of password variations for a given starting password~\cite{pal2019beyond, das2014tangled, wang2016targeted}, primarily with the intention of modeling credential tweaking attacks. Credential tweaking is a targeted attack where the adversary knows the targeted user's credentials for one or more services and attempts to compromise accounts of the same user on other services. 
Different from credential stuffing, here the user's passwords are supposed to be ``tweaked'' versions\footnote{The user can create such password variations to accommodate passwords composition policies of different services.} of the known ones. 
In this direction, Pal~\etal~\cite{pal2019beyond} proposed novel attack/defense techniques for credential tweaking. Both the attack and defense techniques are built on top of a password similarity concept. They model a specific form of semantic similarity by using a supervised dataset of user-password pairs. They assume the \textit{distributional hypothesis} for passwords to be \textbf{true}, and define two passwords to be `similar' if they are often chosen together by users. The proposed attack technique is based on a probabilistic neural model, and it aims to produce tweaked variations of an input password for a tweaking attack. 
Their technique is based on constructing an embedding space 
that is used to estimate the similarity between chosen passwords. This similarity measure is then used to build a ``personalized password strength meter'' that aims to spot the use of a tweaked password by the user at password creation time. In contrast to our password representation, their embedding space does not allow for sampling operation and passwords generation.

\section{Conclusion and future works}
\label{section:conclusion}
We presented a complete paradigm shift in the task of password guessing that is orthogonal to the current research directions. We demonstrated that locality principles imposed by the latent representation of deep generative models open new practical and theoretical possibilities in the field. Based on these properties, we propose two new password guessing frameworks, \ie CPG and DPG. The CPG framework
enables the conditional generation of arbitrarily biased passwords. We empirically demonstrated its inherent advantages with respect to well-established state-of-the-art approaches. In addition, the DPG framework demonstrates that the knowledge from freshly guessed passwords can be successfully generalized and used to mimic the target password distribution. 
More importantly, this guessing technique allows the generation of passwords that are peculiar for the attacked password distribution, and that would require an impractical effort to be guessed by other guessers.

\section*{Availability}
	The code, pre-trained models, and other materials related to our work are publicly available at~\cite{paper_repo}.




\bibliographystyle{plain}
\bibliography{bib.bib}

\appendices

\section{Inducing peculiar password latent organizations via inductive bias}
\label{sec:ind_bias}
\keepme{
Given the absence of precise external bias, the generative models used to learn the latent password representation is free to choose arbitrary spatial arrangements among passwords. In the general case, our generators learn the latent representation that best supports the extremely general generative task imposed during the training. However, this may not be optimal. 
For instance, the latent spaces learned by our technique tend to keep passwords with similar length very close to each other. The reason is that the length of a password is modeled as one of the core explanatory factors~\cite{bengio2013representation} by the latent representation.
As a result, passwords with different lengths are distributed far from each other, which is good for DPG but undesirable in other cases. For instance, it may be better to generate passwords that share specific substrings, but that do not have comparable length.}

\par
\keepme{
Luckily, this type of specialization is possible within our frameworks. Our deep learning approach is highly versatile, and password organizations that present a peculiar feature can be obtained through the injection of inductive bias during the learning process.}
\par
\keepme{
Focusing on the AE (Section~\ref{section:CWAE}), we can indeed induce structure preferences in the latent space organization through regularizations during training. For instance, we can easily reduce the length-based clustering phenomenon described above by acting on the character deletion process used in Section~\ref{section:CWAE}. In the normal case, we learn a latent representation by training the auto-encoder at reconstructing artificially mangled passwords, where each character in the input string is removed with a certain probability. Differently, we can delete a group of $k$ continuous characters given a randomly chosen starting position $i$. For instance, with $k=5$, a password \textit{``jimmy1991''} can become \textit{``jimm*****''} with $i=4$; otherwise \textit{``*****991''} with $i=0$. Intuitively, the generator collects in the same location passwords that share common substrings, regardless of their length. For instance, given the mangled password \textit{``jimmy*****''}, the generator should be able to recover the passwords \textit{``jimmy''}, \textit{``jimmyjimmy''} and \textit{``jimmy123''}, eventually forcing their latent representations to be close to each other.}
\par
\keepme{
As an example, we compare passwords sampled from CWAE trained with different approaches, namely, using the character deletion approach discussed in Section~\ref{section:CWAE} (here, referred to as \textit{Simple}) and using the group deletion approach discussed above (referred to as \textit{Mask}). Table~\ref{tab:sn} reports password sampled around the pivot \textit{``iloveyou1''} for the two CWAEs. Compared to \textit{Simple}, passwords sampled from the \textit{Mask} model tend to have heterogeneous lengths which are arbitrarily different from the one of the pivot.}

\begin{table}[!htbp]
	\vspace{-1em}
	\centering
	\caption{Passwords sampled around the pivot \textit{``iloveyou1''} for two CWAEs trained with different regularization. The same value of $\sigma$ is used for both models.}
	\resizebox{.4\columnwidth}{!}{%
		\label{tab:sn}

		\begin{tabular}{ll}
			\toprule
			\textbf{Simple} & \textbf{Mask} \\
			\midrule
			iloveyou13 &  iloveyou1234 \\
			iloveyou12 &  iloveyou14 \\
			iloveYou1 &  iloveyou12ao \\
			iLoveyou1 &  iloveyou1222 \\
			iloveyou* &  iloveyou17a \\
			Iloveyou1 & iloveyou12arham \\
			iloheyou1 &  iloveyou14om \\
			ilOveyou1 &  iloveyou123o \\
			iloveyou11a &  iloveyou1444 \\
			iloveyou1a & iloveyou12a4mom1 \\

			\bottomrule
		\end{tabular}

	}
\end{table}
\section{Guess generation performance}
\label{sec:bench}
\keepme{
	For the sake of completeness, we report performance analysis of our methods and other probabilistic models, namely, OMEN and PCFG. We exclude FLA~\cite{fla} in this comparison, as to the best of our knowledge, its enumeration algorithm do not produce sorted password in a stream, \ie password must be first pre-computed and then used for the guessing attack. Moreover, during our experiments, generating $10^{10}$ guesses required more than two weeks on an NVIDIA Quadro P6000 and Intel Xeon CPU E5645.\footnote{We failed to deploy FLA's implementation on our most performing machine that we used for the other benchmarks.}}
\par
\keepme{
	Benchmarks are performed without considering the latency induced from the hash computation; only the guess-generation cost is evaluated. However, we write the guesses on disk. For each tested tool, we generate $10^8$ passwords and collect the required execution time. Then, we compute the throughput of the generated guesses per second (g/s). Our implementations come with two options: (1) we allow the generation of duplicate guesses; and (2) we filter repeated guesses from the output stream by using a bloom filter. The former option is useful when fast hash functions are considered while the latter is better suited for slow hash functions. For other tools, we evaluate the implementation using the default settings. Data are reported for both GAN and CWAE models. We perform the tests on an NVIDIA DGX2 machine with NVIDIA V100s (32GB device memory). Table~\ref{table:perf_data} reports the collected data.}
\begin{table}[H]
	\centering
	\caption{Data collected for performance analysis}
	\resizebox{1\columnwidth}{!}{%
		\label{table:perf_data}
	\begin{tabular}{cccccc}
		\toprule
		\textbf{OMEN} & \textbf{PCFG} & \textbf{GAN} & \makecell{\textbf{GAN}\\(filtered)} & \textbf{CWAE} & \makecell{\textbf{CWAE}\\(filtered)}\\ \midrule
		512820 g/s & 114810 g/s & 303951 g/s & 80321 g/s & 237529 g/s & 64197 g/s\\
		\bottomrule
	\end{tabular}
}
\end{table}
\keepme{
Among the tested tools, OMEN is the fastest. It is indeed shipped with a \textit{C} implementation while PCFG and ours are implemented in \textit{python} language. Considering raw hashes (\ie 0 work factor), on the same hardware setup, hashcat can compute $\sim 60000$ \textit{bcrypt} hashes per second on a single GPU. This number becomes 1894.5 Mg/s when a fast hash function, such as \textit{SHA-3 512}, is used. Considering a single iteration of \textit{bcrypt} as a baseline for secure password storing, all the tested tools can saturate the hashing pipeline when bruteforce-aware hashing algorithms are employed, but they fail when fast hash functions are considered. It is important to highlight that the use of fast hash functions is not a secure choice to store passwords.}

\section{Learning the inverse mapping for the GAN model}
\label{section:learn_inverse_mapping}
To fully exploit the properties offered by the learned latent representation of passwords, we need a way to explore the latent space efficiently. Therefore, our primary interest is to understand the relation between the observed data (\ie passwords) and their respective latent representations; in particular, their position within the latent space. A direct way to model this relation is to learn the inverse of the generator function $G^{-1}:\mathbf{X} \rightarrow \mathbf{Z}$. \massimo{GANs, by default, do not need to learn those functions because that} requirement is bypassed by the adversarial training approach. To do so, framework variations~\cite{donahue2016adversarial, dumoulin2016adversarially} or additionally training phases~\cite{luo2017learning} are~required. 
\par
To avoid any source of instability in the original training procedure, we opt to learn the inverse mapping only after the training of the generator is complete. This is accomplished by training a third \textbf{encoder} network $E$ that has an identical architecture as the \textit{critic}, except for the size of the output layer. The network is trained to simultaneously map both the real (\ie data coming from the train-set) and generated (\ie data coming from $G$) data to the latent space. Specifically, the loss function of $E$ is mainly defined as the sum of the two cyclic reconstruction errors over the data space. This is presented in~the~following:
\begin{equation}
\begin{aligned}
L_0 &= \mathbb{E}_z[d(G(z),G(E(G_t(z))))],\\
L_1 &= \mathbb{E}_x[d(x,G(E(x)))].
\end{aligned}
\label{equation:two_cyclic_reconstructions}
\end{equation}
In \massimo{Eq.~(\ref{equation:two_cyclic_reconstructions}), the function $d$ is the cross-entropy whereas $x$ and $z$ are sampled from the train-set and the prior latent distribution, respectively. The variable $t$ in $L_0$ refers to the temperature of the final \textit{softmax} layer of the generator. In Eq.~(\ref{equation:two_cyclic_reconstructions}), we do not specify temperature on a generator notation when it is assumed that it does not} change during the training. The combination of these two reconstruction errors \massimo{aims} at forcing the encoder to learn a general function capable of inverting both the true and generated data correctly. As discussed in Section~\ref{section:model_improvements}, the discrepancy between the representation of the true and generated data (\ie discrete and continuous data) is potentially harmful to the training process. To deal with this issue, we anneal the temperature $t$ in loss term $L_0$ during the training. We do that to collapse slowly the continuous representations of the generated data (\ie the output of the generator) towards the same discrete representation of the real data (\ie coming from the dataset). Next, an additional loss term, shown in Eq.~\ref{equation:additional_loss_term}, is added \massimo{forcing} the encoder to map the data space in a dense zone of the latent space (dense with respect to the prior latent distribution).
\begin{equation}
L_2 = \mathbb{E}_z[d(z, E(G(z)))].
\label{equation:additional_loss_term}
\end{equation}
Our final loss function for $E$ is reported in Eq.~\ref{equation:final_loss_function}. During the encoder training, we use the same train-set that we used to train the generator, but we consider only the unique passwords in this case.
\begin{equation}
L_E = \alpha L_0 + \beta L_1 + \gamma L_2.
\label{equation:final_loss_function}
\end{equation}
The information about the hyper-parameters we used is listed in \tablename{~\ref{table:hyperparameter}}.
\begin{table}[H]
	\centering
	\caption[]{Hyper-parameters used to train our encoder network}
	\resizebox{.4\columnwidth}{!}{%
		\begin{tabular}{ll}
			\toprule
			\multicolumn{1}{l}{\textbf{Hyper-parameter}} & \multicolumn{1}{l}{\textbf{Value}} \\ \midrule
			$\alpha$ & 0.2 \\ 
			$\beta$ & 0.2 \\ 
			$\gamma$ & 0.6 \\ 
			Batch size & 64 \\ 
			Learning rate & 0.001 \\ 
			Optimizer & \textit{Adam} \\ 		
			Temperature decay step & 250000 \\ 
			Temperature limit & 0.1 \\ 
			Temperature scheduler & \textit{polynomial} \\ 
			Train iteration & $3\cdot 10^5$ \\ 
			\bottomrule
		\end{tabular}
	}
	\label{table:hyperparameter}
\end{table}

\section{On the impact of hyper-parameters on DPG}
\label{sec:dpg_h}
In this section, we briefly consider the impact of the two hyper-parameters of DPG over the quality of the attack.\\

\begin{figure}[!htbp]
	\centering
	\includegraphics[trim = 0mm 0mm 0mm 0mm, clip, width=.65\linewidth]{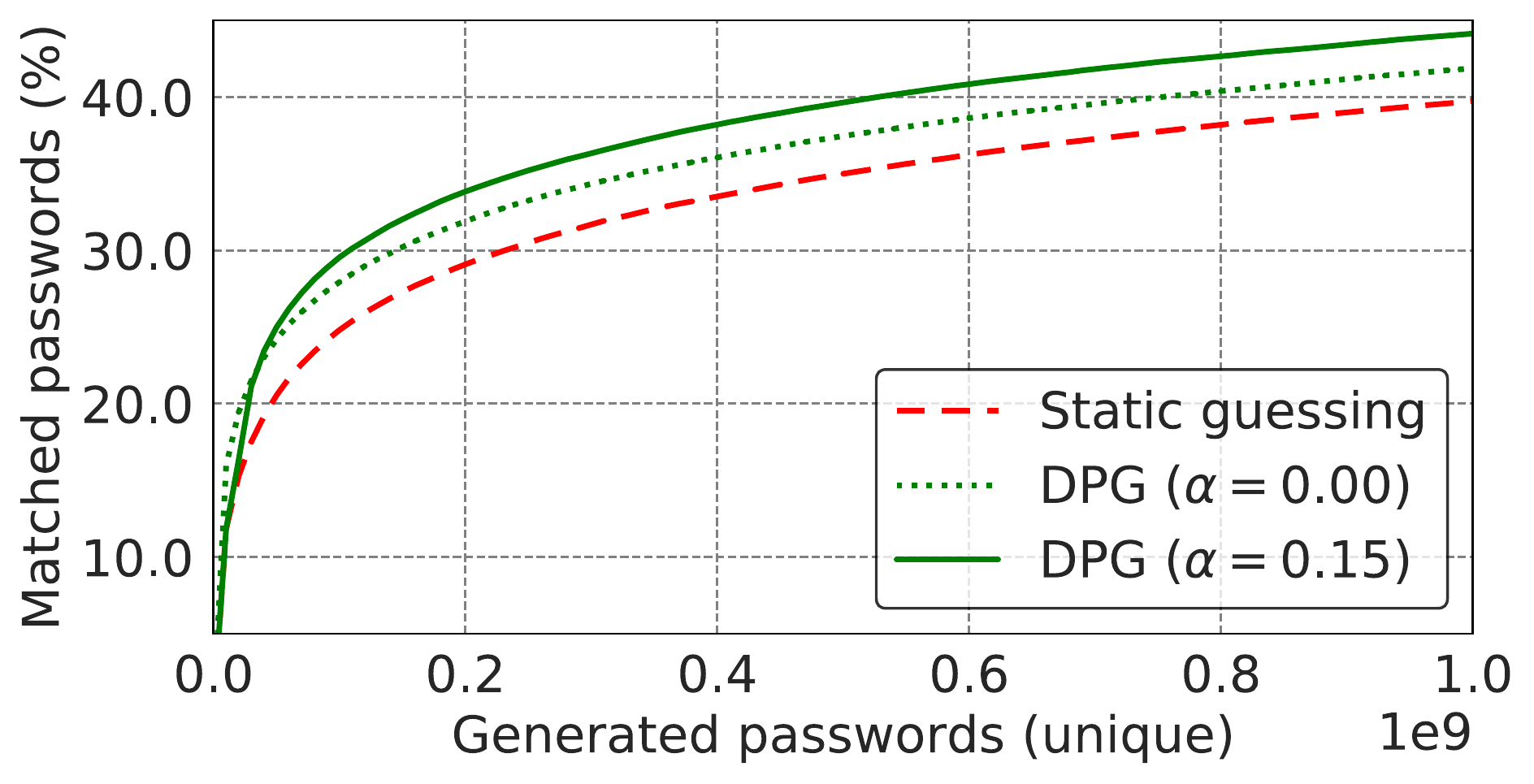}
	\caption[]{The impact of $\alpha$ on the performance of DPG for \textit{phpbb} test-set}
	\label{figure:hot_start_necessity}
\end{figure}

\figurename{~\ref{figure:hot_start_necessity}} depicts a comparison among the static attack, a DPG with $\alpha=15\%$, and a DPG with $\alpha=0\%$ (\ie no hot-start). These results confirm that the absence of hot-start indeed affects and eventually degrades the performance of~DPG.
\par
\begin{figure}[!htbp]
	\centering
	\includegraphics[trim = 0mm 0mm 0mm 0mm, clip, width=.7\linewidth]{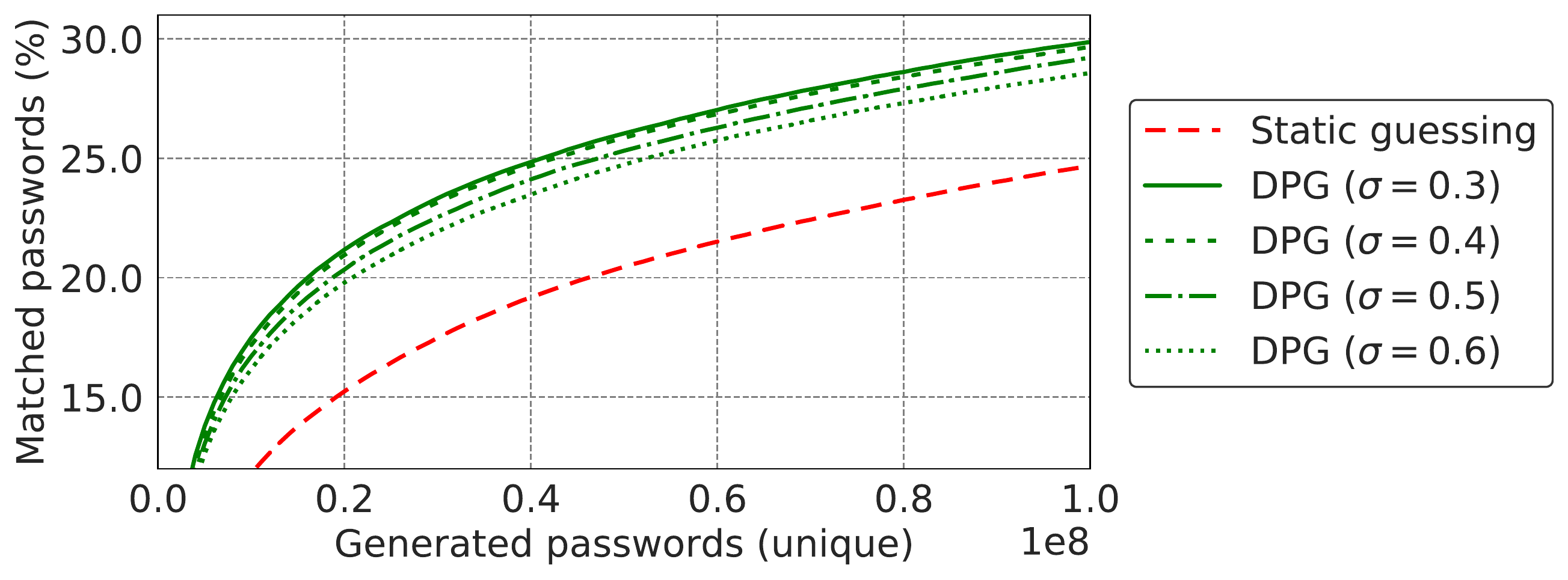}
	\caption[]{The impact of $\sigma$ on the performance of DPG for \textit{phpbb} test-set}
	\label{figure:hot_start_differnt_sigma}
\end{figure}

 \figurename{~\ref{figure:hot_start_differnt_sigma}} depicts the effect of different values of $\sigma$ on the performance of DPG. Smaller values of $\alpha$ yields better overall results. 
This outcome suggests that it is not necessary to sample too far from the dense zones imposed by $Z_i$, and rather a focused exploration of those zones is beneficial. This observation is perfectly coherent with the discussed locality property, giving further support to the speculated ability of the latent space of capturing and translating general features of an entire password distribution in geometric~relations.
\section{CWAE details}
\label{section:CWAEd}
We build the CWAE architecture leveraging the same \textit{resenet-like} structure used for the GAN generator, which is summarized in \tablename{~\ref{table:cwaearch}}. \tablename{~\ref{table:hyperparametercwae}} reports the hyper-parameters used during the training of the model. Here, $\lambda$ is the weight assigned to the latent divergence term in the loss function, \ie Maximum Mean Discrepancy (MMD) for our case. We use a standard \textit{softmax-crossentropy} for the distance measure in the data space. The parameter $\epsilon$ controls the character-deletion probability used during the training (discussed in Section~\ref{section:CWAE}). For more fine-grained information, \ankit{please refer to our project page (see Section~Availability).}
\begin{table}
	\centering
	\caption{CWAE architecture scheme}
	\resizebox{.7\columnwidth}{!}{%
		\begin{tabular}{ll}
			\textbf{Encoder} \\
			\toprule
			\cd{5}{128}{same}{linear} \\ 
			\rb{128}{false} \\ 
			\rb{128}{false} \\ 
			\rb{128}{false} \\ 
			\rb{128}{false} \\ 
			\rb{128}{false} \\ 
			\rb{128}{false} \\ 
			\rb{128}{false} \\ 
			\texttt{Reshape}[-1] \\
			\texttt{FullyConnected}[128, \textit{linear}] \\ \toprule
			\textbf{Decoder} \\ \toprule
			\texttt{FullyConnected}[$\textit{MaxPasswordLength}\cdot128$, \textit{linear}] \\ 
			\texttt{Reshape}[\textit{MaxPasswordLength}, 128] \\ 
			\rb{128}{false} \\
			\rb{128}{false} \\
			\rb{128}{false} \\
			\rb{128}{false} \\
			\rb{128}{false} \\
			\rb{128}{false} \\
			\rb{128}{false} \\
			\cd{1}{AlphabetCardinality}{same}{linear}
		\end{tabular}
	}
	\label{table:cwaearch}
\end{table}
\begin{table}[H]
	\centering
	\caption[]{Hyper-parameters used to train our CWAE}
	\resizebox{.4\columnwidth}{!}{%
		\begin{tabular}{ll}
			\toprule
			\multicolumn{1}{l}{\textbf{Hyper-parameter}} & \multicolumn{1}{l}{\textbf{Value}} \\ \midrule
			$\lambda$ & 8.0 \\  
			Batch size & 256 \\ 
			Learning rate & 0.0001 \\ 
			Optimizer & \textit{Adam} \\ 		
			Train Epochs & $25$ \\ 
			$\epsilon$ & 5.0 \\
			\bottomrule
		\end{tabular}
	}
	\label{table:hyperparametercwae}
\end{table}
\section{Supplementary tables \& figures}
\label{section:supplementary_table_and_figures}
\ankit{Here, we present supplementary data related to our work. \tablename{~\ref{table:template_samples}} lists the samples of password templates and their respective matching passwords.}. Table~\ref{table:linkedin} extends
\keepme{
 Table~\ref{table:peculiar} for the attack on the \textit{LinkedIn} set. We report the guess-numbers for John the Ripper, Hashcat, Markov Model, and PCFG. These value have been obtained via the CMU-PGS~\cite{cmupgs, ur2015measuring}. Note that PGS sets up its models with a different ground-truth; our train-set is just a subset of the one used from PGS.}
\par
\keepme{
In the table, the underscore symbol `$\_$' indicates that the password model failed to match the password. The column `\textit{DPG G.}' reports the guess-number of the dynamic attack. The passwords are sorted using the same criteria used in Table~\ref{table:peculiar}. We report the top $100$ entries.}
\begin{table*}[!htbp]
	\centering
	\caption{Samples of password templates and respective matching passwords.}
	\label{table:template_samples}
	\resizebox{\textwidth}{!}{%
		\begin{tabular}{ccc|ccc|ccc|ccc}
			\toprule
			\multicolumn{3}{c}{\textbf{$T_{\text{common}}$}}  & \multicolumn{3}{c}{\textbf{$T_{\text{uncommon}}$}}  & \multicolumn{3}{c}{\textbf{$T_{\text{rare}}$}}  & \multicolumn{3}{c}{\textbf{$T_{\text{super-rare}}$}}  \\
			\midrule
			\textbf{*a*e*on**} & \textbf{ri***19**} & \textbf{*ol*nd***} & \textbf{Bi**o**1*} & \textbf{**n1**0*0} & \textbf{***dy*78*} & \textbf{a*6*4*0**} & \textbf{**j99*9**} & \textbf{*n****0!!} & \textbf{k*****kbn} & \textbf{**sb*9*8*} & \textbf{*YR**R*U*} \\
			\midrule
			Cameron4\$  & rizal1982  & Colinda23  & BigCorp11  & Mon171050  & sandy@786  & a06142001  & sbj991980  & Qny1960!!  & ktyzhekbn  & mosby9382  & PYR@GR@UP  \\
			cameron64  & rissi1909  & yolanda\#1  & BigFoot13  & Len112080  & sandy6789  & a26042004  & tej991991  & ando140!!  & kgn5*5kbn  & elsb1968!  & MYRATROUT  \\
			CabeZone1  & rimpy1984  & Noland405  & Bishon111  & ben101010  & goody1785  & ab6643014  & Lwj990922  & vny@@00!!  & ktrnhjkbn  & lksbs9080  &   \\
			madelon13  & riana1976  & noland339  & Bigfoot1\#  & chn102030  & cindy2785  & a76645090  & nhj990920  & lnb7280!!  & kbnkbnkbn  & ldsbc9886  &   \\
			Camerone3  & rinni1970  & rolando13  & Bingo2011  & Jan172010  & maddy2789  & a1644104a  & naj999999  & anaid60!!  &   &   &   \\
			cameronq2  & richu1989  & roland589  & Biddoma12  & van102030  & buddy8780  & a26547054  & Slj999999  & @ngel20!!  &   &   &   \\
			makedon24  & rinks1978  & Rolando85  & Bigboy117  & jan152000  & brady1785  & a06042007  & jjj999999  & QnA2010!!  &   &   &   \\
			Kameron76  & rinat1978  & roland006  & Biofoto10  & ten142000  & maddy@786  & a8674600Z  & msj991987  & Annie20!!  &   &   &   \\
			cameron46  & risco1969  & RolandD50  & Biologo12  & jan142000  & sandy7780  & a76042074  & 99j99a99k  & Annie10!!  &   &   &   \\
			Nakedone1  & riken1970  & Jolanda48  & BioComp10  & l4n1n402  & Toodys781  & am68400en  & dej991976  & inusa20!!  &   &   &   \\
			\bottomrule
		\end{tabular}%
	}
\end{table*}

\begin{table}[!t]
	\centering
	\caption{Guess-numbers of the top peculiar password guessed from DPG for \textit{LinkedIn} leak.}
	\label{table:linkedin}
	
	\begin{minipage}[b]{0.48\textwidth}
		
		\resizebox{1\textwidth}{!}{%
		\begin{tabular}{c|c|ccccc}
			\toprule
			\textbf{Guessed P.} & \textbf{DPG G.} & \textbf{JTR G.} & \textbf{Hashcat G.} & \textbf{Markov G.} & \textbf{PCFG G.}\\ \midrule
			o2linkedln & $3.4\cdot 10^{9}$ & \_ & \_ & \_ & \_\\
			w2linkedln & $3.1\cdot 10^{9}$ & \_ & \_ & \_ & \_\\
			ydlinkedln & $3.6\cdot 10^{9}$ & \_ & \_ & \_ & \_\\
			linked6in6 & $4.3\cdot 10^{9}$ & \_ & \_ & \_ & \_\\
			j*linkedln & $4.3\cdot 10^{9}$ & \_ & \_ & \_ & \_\\
			linked!in. & $4.8\cdot 10^{9}$ & \_ & \_ & \_ & $2.1\cdot 10^{14}$\\
			wslinked1n & $4.4\cdot 10^{9}$ & \_ & \_ & \_ & \_\\
			linkedgcin & $2.1\cdot 10^{9}$ & \_ & \_ & \_ & \_\\
			linked6in2 & $5.6\cdot 10^{9}$ & \_ & \_ & \_ & $1.7\cdot 10^{14}$\\
			lslinkedln & $4.5\cdot 10^{9}$ & \_ & \_ & \_ & \_\\
			wtlinkedln & $4.5\cdot 10^{9}$ & \_ & \_ & \_ & \_\\
			9auiirji & $5.5\cdot 10^{9}$ & \_ & \_ & \_ & $7.6\cdot 10^{13}$\\
			g2linkedln & $3.4\cdot 10^{9}$ & \_ & \_ & \_ & \_\\
			cslinkedln & $4.4\cdot 10^{9}$ & \_ & \_ & \_ & \_\\
			ymlinkedln & $5.2\cdot 10^{9}$ & \_ & \_ & \_ & \_\\
			linked4in6 & $4.4\cdot 10^{9}$ & \_ & \_ & \_ & $2.2\cdot 10^{14}$\\
			fvlinkedln & $4.7\cdot 10^{9}$ & \_ & \_ & \_ & \_\\
			jslinkedln & $3.7\cdot 10^{9}$ & \_ & \_ & \_ & \_\\
			jzlinkedln & $5.1\cdot 10^{9}$ & \_ & \_ & \_ & \_\\
			sslinkedln & $4.4\cdot 10^{9}$ & \_ & \_ & \_ & \_\\
			grlinkedln & $4.7\cdot 10^{9}$ & \_ & \_ & \_ & \_\\
			linkedm1x1 & $2.5\cdot 10^{9}$ & \_ & \_ & \_ & \_\\
			svlinked1n & $5.1\cdot 10^{9}$ & \_ & \_ & \_ & \_\\
			m1linkedln & $3.8\cdot 10^{9}$ & \_ & \_ & \_ & \_\\
			linkedi9in & $2.7\cdot 10^{9}$ & \_ & $6.7\cdot 10^{11}$ & \_ & \_\\
			mnlinkedln & $3.7\cdot 10^{9}$ & \_ & \_ & \_ & \_\\
			etlinkedln & $4.9\cdot 10^{9}$ & \_ & \_ & \_ & \_\\
			forc3link & $2.1\cdot 10^{9}$ & $8.6\cdot 10^{11}$ & $1.0\cdot 10^{9}$ & \_ & $4.3\cdot 10^{11}$\\
			5.linkedin & $4.7\cdot 10^{9}$ & \_ & \_ & \_ & $8.8\cdot 10^{11}$\\
			link4rfxa & $4.8\cdot 10^{9}$ & \_ & \_ & \_ & $4.2\cdot 10^{11}$\\
			g0linked1n & $2.5\cdot 10^{9}$ & \_ & \_ & \_ & $1.3\cdot 10^{14}$\\
			linkedm1m1 & $2.9\cdot 10^{9}$ & \_ & $6.7\cdot 10^{11}$ & \_ & \_\\
			56linkedln & $4.6\cdot 10^{9}$ & \_ & \_ & \_ & \_\\
			Rbnoi076 & $2.0\cdot 10^{9}$ & \_ & \_ & \_ & $4.2\cdot 10^{13}$\\
			linkedtgin & $1.9\cdot 10^{9}$ & \_ & \_ & \_ & \_\\
			linked8in4 & $5.6\cdot 10^{9}$ & \_ & \_ & \_ & $2.0\cdot 10^{14}$\\
			linked!in1 & $4.4\cdot 10^{9}$ & \_ & \_ & \_ & $4.1\cdot 10^{13}$\\
			imlindedin & $4.9\cdot 10^{9}$ & \_ & \_ & \_ & \_\\
			linkedkbin & $2.9\cdot 10^{9}$ & \_ & \_ & \_ & \_\\
			linked9in6 & $4.2\cdot 10^{9}$ & \_ & \_ & \_ & \_\\
			htlinkedln & $4.8\cdot 10^{9}$ & \_ & \_ & \_ & \_\\
			golinkedln & $5.2\cdot 10^{9}$ & \_ & \_ & \_ & \_\\
			ozlinkedln & $5.1\cdot 10^{9}$ & \_ & \_ & \_ & \_\\
			o.linkedin & $4.3\cdot 10^{9}$ & \_ & $6.7\cdot 10^{11}$ & \_ & $6.3\cdot 10^{11}$\\
			linkedwcz & $2.9\cdot 10^{9}$ & \_ & \_ & \_ & $2.9\cdot 10^{13}$\\
			linked\_iin & $5.0\cdot 10^{9}$ & \_ & \_ & \_ & $4.6\cdot 10^{13}$\\
			linkedrcin & $1.6\cdot 10^{9}$ & \_ & \_ & \_ & \_\\
			42linkedln & $4.5\cdot 10^{9}$ & \_ & \_ & \_ & \_\\
			linkedcmw4 & $3.1\cdot 10^{9}$ & \_ & \_ & \_ & \_\\
			mmlinkedln & $3.7\cdot 10^{9}$ & \_ & \_ & \_ & \_\\
		\end{tabular}%
		}

	\end{minipage}\begin{minipage}[b]{0.5\textwidth}
		\centering
		\begin{tabular}{c|c|ccccc}
		2xrilidi & $5.1\cdot 10^{9}$ & \_ & \_ & \_ & $1.8\cdot 10^{13}$\\
		dslinkedln & $4.3\cdot 10^{9}$ & \_ & \_ & \_ & \_\\
		linkedtdin & $2.1\cdot 10^{9}$ & \_ & \_ & \_ & \_\\
		linked1.in & $3.2\cdot 10^{9}$ & \_ & \_ & \_ & $2.0\cdot 10^{14}$\\
		linked4in2 & $4.4\cdot 10^{9}$ & \_ & \_ & \_ & $8.6\cdot 10^{13}$\\
		linked4in4 & $4.1\cdot 10^{9}$ & \_ & $6.7\cdot 10^{11}$ & \_ & $1.0\cdot 10^{14}$\\
		linked.4in & $3.2\cdot 10^{9}$ & \_ & \_ & \_ & $2.2\cdot 10^{14}$\\
		pdlinkedln & $4.2\cdot 10^{9}$ & \_ & \_ & \_ & \_\\
		oklinkedln & $5.2\cdot 10^{9}$ & \_ & \_ & \_ & \_\\
		Or2nge47 & $1.7\cdot 10^{9}$ & \_ & $1.1\cdot 10^{11}$ & \_ & \_\\
		z1linkedin & $4.3\cdot 10^{9}$ & \_ & $6.7\cdot 10^{11}$ & \_ & $8.8\cdot 10^{10}$\\
		linkednxin & $1.4\cdot 10^{9}$ & \_ & \_ & \_ & \_\\
		53linkedln & $4.9\cdot 10^{9}$ & \_ & \_ & \_ & \_\\
		linkedctq & $2.8\cdot 10^{9}$ & \_ & \_ & \_ & $2.7\cdot 10^{13}$\\
		odlinkedln & $3.5\cdot 10^{9}$ & \_ & \_ & \_ & \_\\
		omlinkedln & $3.9\cdot 10^{9}$ & \_ & \_ & \_ & \_\\
		eu293634r & $2.3\cdot 10^{9}$ & \_ & \_ & \_ & $5.4\cdot 10^{13}$\\
		hklinked1n & $5.0\cdot 10^{9}$ & \_ & \_ & \_ & \_\\
		linkedfsin & $2.0\cdot 10^{9}$ & \_ & $6.7\cdot 10^{11}$ & \_ & \_\\
		lf00garl & $4.3\cdot 10^{9}$ & \_ & $8.8\cdot 10^{11}$ & \_ & $2.6\cdot 10^{12}$\\
		y9linkedin & $5.3\cdot 10^{9}$ & \_ & \_ & \_ & $9.2\cdot 10^{11}$\\
		linked87ln & $2.5\cdot 10^{9}$ & \_ & \_ & \_ & $1.1\cdot 10^{14}$\\
		linked544y & $2.8\cdot 10^{9}$ & \_ & \_ & \_ & $8.9\cdot 10^{13}$\\
		xbCA0N & $1.9\cdot 10^{9}$ & \_ & \_ & \_ & \_\\
		linktebow & $9.1\cdot 10^{8}$ & \_ & \_ & \_ & $2.8\cdot 10^{13}$\\
		y2linkedin & $4.5\cdot 10^{9}$ & \_ & \_ & \_ & $4.1\cdot 10^{11}$\\
		linkedmiam & $3.7\cdot 10^{9}$ & \_ & \_ & \_ & \_\\
		73linkedln & $4.1\cdot 10^{9}$ & \_ & \_ & \_ & \_\\
		alasEN00 & $4.4\cdot 10^{8}$ & \_ & \_ & \_ & $9.3\cdot 10^{11}$\\
		h9linkedin & $4.5\cdot 10^{9}$ & \_ & \_ & \_ & $7.9\cdot 10^{11}$\\
		linkedkbl & $1.9\cdot 10^{9}$ & \_ & $6.7\cdot 10^{11}$ & \_ & $2.6\cdot 10^{13}$\\
		T8wtas00 & $5.7\cdot 10^{8}$ & \_ & \_ & \_ & $5.3\cdot 10^{13}$\\
		linkedw3s & $4.0\cdot 10^{9}$ & \_ & \_ & \_ & \_\\
		44linkedln & $4.7\cdot 10^{9}$ & \_ & \_ & \_ & \_\\
		unpceddi & $5.1\cdot 10^{9}$ & \_ & \_ & \_ & $2.4\cdot 10^{11}$\\
		linkedwge & $1.1\cdot 10^{9}$ & \_ & $6.7\cdot 10^{11}$ & \_ & $2.5\cdot 10^{13}$\\
		linked39in & $2.3\cdot 10^{9}$ & \_ & \_ & \_ & $1.1\cdot 10^{14}$\\
		linked99ln & $2.3\cdot 10^{9}$ & \_ & \_ & \_ & $8.7\cdot 10^{13}$\\
		linke14din & $3.1\cdot 10^{9}$ & \_ & \_ & \_ & $1.8\cdot 10^{13}$\\
		gxlinkedln & $3.3\cdot 10^{9}$ & \_ & \_ & \_ & \_\\
		linkedkpin & $1.2\cdot 10^{9}$ & \_ & \_ & \_ & \_\\
		gonulelif & $2.5\cdot 10^{9}$ & \_ & \_ & \_ & $2.5\cdot 10^{13}$\\
		linkedcsun & $2.1\cdot 10^{9}$ & \_ & \_ & \_ & \_\\
		lclinkedln & $4.4\cdot 10^{9}$ & \_ & \_ & \_ & \_\\
		9.linkedin & $4.7\cdot 10^{9}$ & \_ & \_ & \_ & $1.1\cdot 10^{12}$\\
		grswbon3 & $2.5\cdot 10^{9}$ & \_ & \_ & \_ & $2.7\cdot 10^{12}$\\
		snlinkedln & $4.8\cdot 10^{9}$ & \_ & \_ & \_ & \_\\
			\bottomrule
		\end{tabular}%
	
	\end{minipage}
\end{table}

\end{document}